\documentclass[floatfix,superscriptaddress,showpacs,amssymb,10pt,aps,prd,reprint,longbibliography]{revtex4-1}

\usepackage{graphicx,epsfig,amssymb} 
\usepackage{amsmath,amsfonts, times}
\usepackage{bm} 

\usepackage[linktocpage,colorlinks]{hyperref}
\usepackage[caption=false]{subfig}
\usepackage[usenames]{color}     
\usepackage{natbib}
\usepackage{soul}

\usepackage[utf8]{inputenc}

\usepackage{float}

\definecolor{coolblack}{rgb}{0.0, 0.18, 0.39}
\definecolor{darkred}{rgb}{0.5,0,0}
\definecolor{darkgreen}{rgb}{0,0.5,0}
\definecolor{darkblue}{rgb}{0,0,0.5}
\definecolor{lapislazuli}{rgb}{0.15, 0.38, 0.61}
\definecolor{venetianred}{rgb}{0.78, 0.03, 0.08}
\definecolor{bleudefrance}{rgb}{0.19, 0.55, 0.91}
\definecolor{dogwoodrose}{rgb}{0.84, 0.09, 0.41}
\definecolor{purple}{rgb}{0.58, 0, 0.82}
\definecolor{green}{RGB}{0, 128, 0}
\hypersetup{colorlinks=true, citecolor=darkgreen, linkcolor=darkblue, 
	urlcolor = blue}

\def\be{\begin{equation}}
	\def\ee{\end{equation}}

\newcommand{\dd}{\mathrm{d}}
\newcommand{\expe}{\operatorname{e}}

\begin{document}
	
	\author{Qian Li}
	\address{School of Physics and Technology, Wuhan University, Wuhan, 430072, China}
	\author{Qianchuan Wang}
	\address{School of Physics and Technology, Wuhan University, Wuhan, 430072, China}
	
	\author{Junji Jia}
	\email[Corresponding author:~]{junjijia@whu.edu.cn}
	\address{Department of Astronomy $\&$ MOE Key Laboratory of Artificial Micro- and Nano-structures, School of Physics and Technology, Wuhan University, Wuhan, 430072, China}
	\title{\Large On-axis absorption and scattering of charged massive scalar waves by Kerr-Newman black-bounce spacetime}
	
	\begin{abstract}
		We investigate the absorption and scattering of charged massive scalar waves by the Kerr-Newman black-bounce spacetime when the waves are incident along the rotation axis. We calculate the geometrical and glory scattering cross sections using the classical analytical method, and the corresponding absorption and scattering cross sections using the partial wave method, showing that they are in excellent agreement. Our findings indicate that a faster (slower) rotating spacetime or a more repulsive (attractive) electric force tends to reduce (increase) the absorption cross section and results in larger (smaller) angular widths of the scattered wave oscillations. We find that the rotation parameter exerts a suppressive influence on superradiance, which contrasts with the enhancing effect of the repulsive electric force. It is worth mentioning that the regularization parameter $k$ is found to modify the absorption or scattering cross sections only weakly, but can cause a noticeable reduction of superradiance. To further clarify the role of the parameters in superradiance, we study the energy extraction efficiency in the electric Penrose process. For particles moving along the rotation axis, we find that the influence of the parameters ($a, q, k$) on this efficiency is consistent with their effects on superradiance. We also discuss potential astrophysical applications, showing that particles in this process could be accelerated to ultrahigh energies in realistic environments, and could therefore be used to constrain black hole parameters. For the effect of field mass, it is found that a heavier scalar field leads to a larger absorption cross section and a wider interference fringe of the differential scattering cross section.  When superradiance happens, i.e., the absorption cross section becomes negative, it is also found that the differential scattering cross section only changes smoothly, with no apparent qualitative feature showing up.
		
	\end{abstract}
	\keywords{charged scalar wave, Kerr-Newman black-bounce spacetime, glory scattering, scattering cross section, superradiance}
	
	\maketitle
	\section{Introduction}
	
	Although general relativity has been tested with remarkable success, such as predicting the existence of black holes (BH), it still suffers from the problem of singularities at the BH centers.   It is thought that a theory of quantum gravity could circumvent the problem.  Owing to the lack of a complete quantum theory of gravity, one of the methods to solve the singularity in BHs is to construct regular BHs with finite curvature at their cores. The first regular BH, proposed by Bardeen \cite{Bardeen:1968}, was reinterpreted by Ay\'on-Beato and Garc\'ia \cite{Ayon-Beato:2000mjt} as a magnetic solution that couples Einstein's gravity to nonlinear electrodynamics.  Subsequently, on the basis of Bardeen's idea, numerous studies have focused on proposing other (rotating) regular BHs \cite{Ayon-Beato:1998hmi,Bronnikov:2000vy,Dymnikova:2004zc,Hayward:2005gi,Bambi:2013ufa,Balart:2014cga,Toshmatov:2014nya,Dymnikova:2015hka}. Additionally, Simpson and Visser \cite{Simpson:2018tsi} introduced a new type of regular BH, known as a ``black-bounce" BH, by replacing the coordinate $r$ with $r_k\equiv\sqrt{r^2+k^2}$ in the Schwarzschild BH. Using the Newman-Janis algorithm, Mazza \textit{et al.} in Ref. \cite{Mazza:2021rgq}  generalized the black-bounce (Simpson-Visser) spacetime to a rotating case. Franzin \textit{et al.} \cite{Franzin:2021vnj} introduced a Reissner-Nordstr$\ddot{\rm{o}}$m (RN) and Kerr-Newman (KN) black-bounce spacetimes by applying the Simpson-Visser method.   It is worth noting that in general relativity, the black-bounce spacetime cannot be explained by considering a scalar field or nonlinear electrodynamics alone \cite{Bronnikov:2021uta}. 	
	
	For many years, the study of particles or fields in the vicinity of  BHs has been a research topic, as the behavior of these particles or fields can provide valuable insights into the properties of BHs.  From this perspective, it is interesting and necessary to investigate the absorption and scattering of fields with different spins by BHs.   In particular, it was shown \cite{Crispino:2009xt} that it may be possible to identify quiescent BHs indirectly by the ``fingerprints” they leave on incident radiation, where fingerprints refers to the unique characteristics of a BH's response to external perturbations. Such fingerprints include interference patterns (i.e., scattering cross sections) generated when BHs are illuminated by long-lasting planar radiation. Furthermore, BHs with the same asymptotic properties but different near-horizon geometries can exhibit significant differences in the scattering cross sections of external waves, which may have important astrophysical implications, for example, in testing the no-hair hypothesis \cite{Gussmann:2016mkp}.  In fact, over the past few decades, a lot of effort has been invested in the calculation of absorption and scattering of fields with all kinds of spins by various BHs, such as scalar ($s=0$) \cite{Sanchez:1977si,Crispino:2007zz,Crispino:2009ki,Chen:2011jgd,Macedo:2013afa,Macedo:2014uga,Leite:2016hws,Anacleto:2019tdj,Anacleto:2020lel,Li:2021epb,Li:2022wzi,Sun:2023woa,Wan:2022vcp,Jung:2004yh,Benone:2014qaa}, Dirac ($s=1/2$) \cite{Dolan:2006vj,Sporea:2017zxe}, electromagnetic ($s=1$) \cite{Crispino:2007qw,Crispino:2008zz,Crispino:2009xt,Leite:2017zyb,Leite:2018mon,deOliveira:2019tlk}, and gravitational  ($s=2$)  \cite{Dolan:2008kf,Crispino:2015gua} fields.   	However, relatively limited attention has been given to the absorption and, particularly, the scattering of a charged scalar field by charged BHs \cite{Benone:2015bst,Benone:2019all,Richarte:2021fbi,dePaula:2024xnd}. As is well known in this field, when a test field with integer spin interacts with rotating BHs, under certain conditions, one finds that rotating BHs will amplify the scattering waves \cite{Misner:1972kx}, known as (rotational) superradiance.  Besides, superradiance also occurs in the process where a charged scalar field is scattered by a charged BH.   In this superradiant regime, the energy of the BHs is transferred to the test field as the field extracts mass, charge, and angular momentum from the BH \cite{Brito:2015oca}. This leads to various interesting phenomena, including the negative absorption cross section \cite{Benone:2015bst,Benone:2019all,dePaula:2024xnd}.  However, when investigating the scattering of neutral scalar waves by a  Kerr BH, Glampedakis and Andersson \cite{Glampedakis:2001cx} found that the effect of superradiance on the wave scattering is negligible. To our knowledge, the influence of superradiance caused by a charged rotating BH interacting with a charged scalar field on the scattering cross section remained an open question. 
	
	To investigate whether black-bounce spacetimes can modify existing features of field scattering, researchers have conducted analyses of these spacetimes by studying the behavior of particles or fields around them \cite{Churilova:2019cyt, Nascimento:2020ime,Guerrero:2021ues, Guo:2021wid, Murodov:2024oym, Lima:2020auu,Franzin:2022iai, LimaJunior:2022zvu,Calza:2022ioe,Ghosh:2022mka}.  
	In this paper, we primarily investigate the absorption and scattering of a charged massive scalar field propagating along the rotation axis in the KN black-bounce spacetime, aiming to investigate the effects of electromagnetic interactions on the corresponding cross sections. Additionally, we thoroughly examine how various spacetime and field parameters, particularly the spacetime spin $a$ and regularization parameter $k$, influence superradiance during this scattering process. To better understand the physics of superradiance, we introduce and briefly discuss the electric Penrose process (EPP) in this spacetime. We further consider the potential astrophysical relevance of the EPP in this spacetime. Our findings indicate that, during superradiance, the on-axis incidence leads to enhanced superradiance intensity in slower-rotating spacetimes, which contrasts with the behavior observed for the equatorial incidence in KN BHs \cite{Benone:2019all}. This trend is consistent with the role of the parameter $a$ in the EPP, where, for particles moving along the rotation axis, a larger spin suppresses the energy extraction efficiency.  Furthermore, stronger repulsive electric fields are found to further amplify superradiance intensity.  Notably, we show that the regularization parameter $k$ suppresses superradiance, which is consistent with the results of Ref. \cite{Franzin:2022iai}. In our work, however, the reflected wave extracts Coulomb energy rather than rotational energy from the BH. Thus, it is the first time that the effect of the regularization parameter $k$ on superradiance is studied in such a context. We also report that, for the differential scattering cross sections, no qualitative changes—such as a flip in sign, as observed in the absorption cross section—are detected when superradiance occurred.

	The structure of this work is as follows: In Sec. \ref{sec:metric}, we introduce the KN black-bounce spacetime and the corresponding four-potential. In Sec. \ref{sec:geodesic}, we first obtain the geodesic equations and carry out a basic analysis of the critical surface, after which we investigate the geodesic absorption and scattering phenomena, as well as the glory effect associated with charged and massive particles. In Sec. \ref{sec:partial wave}, we provide a concise overview of the partial wave method, which is employed to derive the expressions for absorption and scattering cross sections in this method. In Sec. \ref{sec:numerical results}, we present and compare the absorption and scattering cross sections obtained using different methods and discuss the influence of various parameters on them. Section \ref{sec:superradiance} is devoted to discussing superradiance. Section \ref{sec:conclusion} concludes the work. Throughout this work, the natural units ($G=c=\hbar=4\pi\varepsilon_0=1$) and the metric signature ($-,+,+,+$) are adopted.

	\section{KN black-bounce spacetime}
	\label{sec:metric}
	The  KN black-bounce metric in Boyer-Lindquist coordinates  can be written as \cite{Franzin:2021vnj}
	\begin{align}\label{eq:spacetime}
		\dd s^2=&-\frac{\Delta}{\rho^2}\left(a\sin^2 \theta  \dd \phi-\dd t\right)^2 + \rho^2 \left(\frac{\dd r^2}{\Delta}+\dd \theta^2\right) \nonumber \\  &+ \frac{\sin^2\theta}{\rho^2}\left[\left(r^2+k^2+a^2\right)\dd \phi -a  \dd t\right]^2
	\end{align}
	with
	\begin{align}\label{eq:rho}
		\rho^2&=r^2+ k^2+a^2 \cos^2\theta, \\  
		\Delta&= r^2+k^2+a^2-2M\sqrt{r^2+ k^2}+Q^2,
	\end{align}
	where $M$ is the spacetime mass,  $a$ is the spin angular momentum per unit mass of the spacetime, $Q$ denotes the charge of spacetime and $k$ stands for the regularization parameter that avoids the existence of the central singularity. It is obvious that when $k=0$, this spacetime will reduce to the KN spacetime. The  event horizon radius is the root of $\Delta$, given by
	\begin{align}\label{eq:rh}
		r_h=\sqrt{\left[M+\sqrt{M^2-(a^2+Q^2)}\right]^2- k^2},
	\end{align}
	and the parameters are such in this work that the inequalities $M^2-(a^2+Q^2)>0$ and $M+\sqrt{M^2-(a^2+Q^2)}> k$ are always satisfied.
	The electromagnetic potential of this spacetime is described by the four potential
	\begin{align}\label{eq:elec potential}
		A_{\alpha}=-\frac{Q\sqrt{r^2+ k^2}}{\rho^2}\left(1,0,0,-a \sin^{2}{\theta}\right).
	\end{align}

	\section{Geodesic analysis}\label{sec:geodesic}
	
	\subsection{Geodesic scattering}\label{subsec:Geodesic}	
	In this section, we investigate the motion of the charged massive particle in the KN black-bounce spacetime in order to better understand the geodesic scattering. The equations of motion can be derived from the Hamilton-Jacobi equation \cite{Carter:1968rr}
	\begin{align}\label{eq:Hamilton-Jacobi}
		2\frac{\partial \mathcal{S}}{\partial \lambda}=g^{\alpha\beta}\big(\frac{\partial \mathcal{S}}{\partial x^{\alpha}}-q A_{\alpha}\big)\big(\frac{\partial \mathcal{S}}{\partial x^{\beta}}-q A_{\beta}\big),
	\end{align}
	where $\mathcal{S}$, $\lambda = \tau/\mu$, and $q$ stand for the action of the test particle, the affine parameter of the motion (with $\tau$ being the proper time and $\mu$ being the mass of the test particle), and the charge of the test particle, respectively.	Due to the axial and stationary symmetries of the background spacetime, associated with  Killing vectors, we have two conserved quantities, namely the energy $E$ and angular momentum $L_z$. It should be noted that in the semiclassical limit, these two quantities are linked to $\omega$ and $l+1/2$, where $\omega$ is the frequency of the incident wave and $l$ is the angular quantum number. Therefore, due to these conserved quantities, the action can always be written in the form	
	\begin{align}\label{eq:Hamilton-solution}
		\mathcal{S}=-\frac{1}{2}\mu^2 \lambda- E t+L_z \phi+ \mathcal{S}_{r}(r)+ \mathcal{S}_{\theta}(\theta).
	\end{align}

	Substituting Eqs. \eqref{eq:spacetime}  and \eqref{eq:Hamilton-solution} into  Eq. \eqref{eq:Hamilton-Jacobi}, we obtain two differential equations
	\begin{subequations}
		\begin{align}
			\Delta^2 \left(\frac{\dd  \mathcal{S}_r}{\dd r} \right)^2=&\left[E\left(r^2+k^2+a^2\right)-aL_z-qQ\sqrt{r^2+ k^2} \right]^2   \nonumber \\ 
			-&\Delta\left[\mathcal{K}+\left(L_z-aE\right)^2+\mu^2\left(r^2+ k^2\right)\right],\\
			\left(\frac{\dd  \mathcal{S}_{\theta}}{\dd  \theta} \right)^2 =& \mathcal{K}+a^2\left(E^2-\mu^2\right)\cos^2\theta-L_z^2 \cot^{2}\theta,
		\end{align}\label{eq:S_R}
	\end{subequations}
	where $\mathcal{K}=K-(L_z-a E)^2$ is known as the Carter constant with $K$ being the separation constant.
	
	To derive the equations of motion, we have to introduce  the canonical momenta $P_{\alpha}$  such that 
	\begin{align}\label{eq:S_canonical}
		P_{\alpha}=\frac{\partial \mathcal{S}}{\partial x^{\alpha}}=  g_{\alpha\beta}\frac{\dd  x^{\alpha}}{\dd  \lambda}+ qA_{\alpha},
	\end{align}
	and we have $P_{t}=-E$ as well as $P_{\phi}=L_z$.
	Substituting the solutions of Eqs. \eqref{eq:Hamilton-solution}  and  \eqref{eq:S_R} into  Eq. \eqref{eq:S_canonical}, we obtain the equations of motion 
	\begin{widetext}
		\begin{subequations}
			\begin{align}
				\centering
				\rho^2_E  \frac{\dd  t}{\dd  \lambda}&=\frac{\left(r^2+k^2+a^2\right)\left[
					\left(r^2+k^2+a^2\right)-a\xi -qQ\sqrt{\left(r^2+k^2\right)}/E \right]}{\Delta}-a\left(a \sin^2\theta-\xi\right), \label{eq:Motion_1}
				\\
				\bigg(\rho^2_E \frac{\dd  r}{\dd  \lambda}\bigg)^2&= \left[\left(r^2+k^2+a^2\right)-a\xi-\frac{qQ\sqrt{\left(r^2+k^2\right)}}{E} \right]^2-\Delta\left[\eta+\left(\xi-a\right)^2+\left(1-v^2\right)\left(r^2+ k^2\right)\right]\equiv R(r),\label{eq:Motion_2} \\
				\rho^2_E 	\frac{\dd  \phi}{\dd  \lambda} &=  \frac{a\left[\left(r^2+k^2+a^2\right)-a\xi-qQ\sqrt{r^2+ k^2}/E \right]}{\Delta}-\left(a-\xi\csc^2\theta\right),\label{eq:Motion_3}\\
				\bigg(\rho^2_E \frac{\dd  \theta}{\dd  \lambda}\bigg)^2&=\eta+a^2 v^2\cos^2\theta-\xi^2 \cot^{2}\theta \equiv \Theta(\theta) \label{eq:Motion_4},
			\end{align}
		\end{subequations}
	\end{widetext}
    where $\rho^2_E = \rho^2 / E$, $\xi = L_z / E$, $\eta = \mathcal{K} / E^2$, and $v = \sqrt{1 - \mu^2 / E^2}$. We also denote the right-hand sides of the radial Eqs. \eqref{eq:Motion_2} and \eqref{eq:Motion_4} as functions $R(r)$ and $\Theta(\theta)$, respectively, to simplify the latter notation.

	Considering the polar orbit (on-axis incidence) with $\xi=0$, the impact parameter is defined as \cite{Macedo:2013afa,Dolan:2010wr} 
	\begin{align}\label{eq:b}
		b=\sqrt{\frac{\eta}{v^2}+a^2}. 
	\end{align}
	Substituting $\xi=0$ and $\eta$ from the above equation into $R(r)$, it becomes explicitly dependent on both $r$ and $b$. Solving the critical conditions $R(r_c,b_c)=0$ and $R'(r_c,b_c)=0$, the critical radius $r_c$ and the critical impact parameter $b_c$ of the trajectory can be obtained numerically. The geometric cross section, which is the absorption cross section in the high-frequency limit, is then directly related to $b_c$ by
	\begin{align}\label{eq:gcs}
		\sigma_{\text{gcs}}=\pi b_c^2.
	\end{align}
	
	For a better understanding of the effects of various parameters on this cross section and the absorption cross section obtained using the partial wave method in Sec. \ref{sec:partial wave}, it is worth studying the dependence of $b_c$ as well as $r_c$ on the main parameters such as $k,\,\omega$ (or equivalently $E$) and $\mu$.  First, noting that all $r^2$ in function $R(r)$ are combined with a $k^2$ and vice versa, it is immediately clear that the $r_c$ here is related to the critical radius of KN spacetime (with $k=0$) by a simple relation $r_c^{\text{KN}}=\sqrt{r_c^2+k^2}$, just as the relation between the critical radius of the neutral Kerr spacetime and its corresponding black-bounce spacetime. Moreover, this also implies that the critical $b_c$ will not depend on the value of the regularization parameter $k$ since any of its variations will be canceled by the corresponding variation in $r_c$. In other words, the KN and KN black-bounce spacetimes will have the same $b_c$ regardless of the value of $k$, as found for the Kerr black-bounce BH in Ref. \cite{Lima:2021las}.  
	
	\begin{figure} 
		\centering
		\includegraphics[width=0.48 \textwidth]{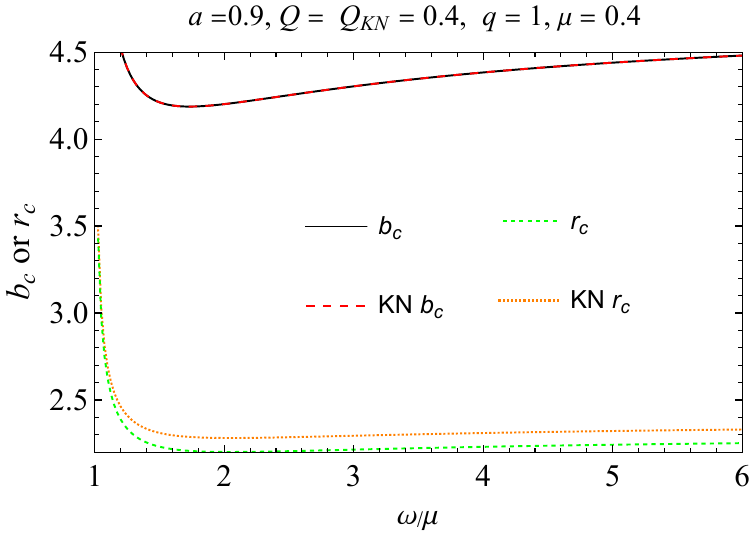}
		\includegraphics[width=0.48 \textwidth]{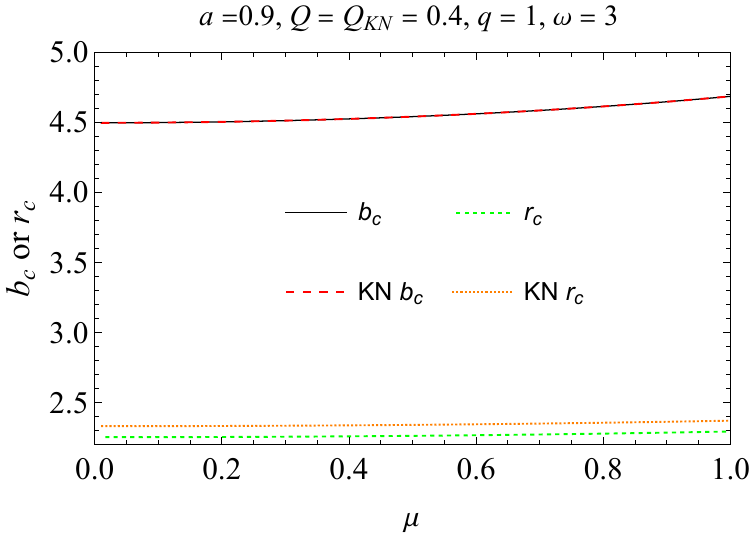}
		\caption{	
			$r_c$ and $b_c$ of a charged particle as functions of $\omega$ (top) and $\mu$ (bottom) in KN black-bounce spacetime with $k=0.6$ and KN BH with fixing $a=0.9$, $Q=0.4$ and $q=1$.}	\label{fig:rc-bc}
	\end{figure}
	In our previous work \cite{Li:2024xyu},  we found that the $b_c$ and $r_c$ of a charged massive particle in the RN BH spacetime initially decrease and then increase with increasing particle energy, but keep a tendency to increase with increasing particle mass.  To investigate whether similar properties of $b_c$ and $r_c$ exist for KN black-bounce spacetime, we plot the $b_c$ and $r_c$ as functions of $\omega$ (top plot) and $\mu$ (bottom plot) for a KN black-bounce spacetime and KN BH in Fig. \ref{fig:rc-bc}.  It should also be noted that unless otherwise specified, we set $M=1$ throughout the paper so that the unit of spacetime parameters is $M$, while the unit of field parameters is $M^{-1}$. In Fig. \ref{fig:rc-bc} (top), we can see that $b_c$ of both KN black-bounce spacetime and KN BH overlaps. However, the presence of repulsive electromagnetic interaction modifies the original decreasing trend observed in the neutral case as the particle energy $\omega$ increases, leading to an initial decrease followed by an increase in both $r_c$ and $b_c$. This qualitative behavior with increasing $\omega$ is thus analogous to the case of the RN spacetime \cite{Li:2024xyu}. 
	For the effect of particle mass $\mu$ on $b_c$ and $r_c$, from the bottom plot of Fig. \ref{fig:rc-bc}, we observe that all of $b_c$ and $r_c$ in both KN black-bounce and KN BH spacetimes grow as particle mass increases while the energy remains fixed. This is in alignment with the simple intuition that a heavier particle with a lower incoming velocity is more likely to be captured even at larger impact parameters.

	From the radial and angular equations \eqref{eq:Motion_2} and  \eqref{eq:Motion_4}, together with the on-axis condition $\xi=0$ and Eq. \eqref{eq:b}, we get the following orbital equation
	\begin{align}\label{eq:orbital equation}
		\bigg(\frac{\dd r}{\dd \theta}\bigg)^2=  \frac{R(r)}{\Theta_b},
	\end{align}
	where $R(r)$ was given in Eq. \eqref{eq:Motion_2} and 
	$\Theta_b=b^2 v^2-a^2 v^2 \sin^2\theta$.
	Using this equation, the deflection angle $\Delta\theta$ is defined as in the upper limit of the following integral \cite{Dolan:2008kf,Chandrasekhar}
	\begin{align} \label{eq:b_theta}
		\int_{0}^{(\Delta\theta+\pi)/2} \frac{\dd \theta}{\sqrt{\Theta_b}}=-\int_{\infty}^{r_0} \frac{\dd r}{\sqrt{R(r)} },
	\end{align}	
	where  $r_0$ is the radius of the turning point of the trajectory, determined as the largest real root of $R(r)=0$. This equality establishes a relation between the deflection angle $\Delta\theta$ and the impact parameter $b$, which we will denote as $b=b(\Delta\theta)$. The classical scattering cross section  can then be calculated as
	\begin{align}\label{eq:cdscs}
		\frac{\dd \sigma}{\dd \Omega}=\sum_{N} \frac{b}{\sin\theta}\bigg|\frac{\dd b}{\dd \Theta}\bigg|, 
	\end{align}
	where $\theta$, the scattering angle, is related to the deflection angle $\Delta\theta$ by $\theta = \left|\Delta\theta - 2N\pi\right|$ with $N=0,\,1,\,2,\,3,\cdots$. Here, $N$ stands for the number of loops that the particle moves around the gravitational center. This cross section will be used in Sec. \ref{sec:numerical results} to compare with the corresponding scattering cross section obtained using partial wave analysis. 
	
	\subsection{Glory scattering } \label{subsec:Glory}	
	We notice that the classical scattering cross section describes the feature of the motion of test particles but fails to account for wave effects, such as interference effects at large scattering angles. Therefore, the glory scattering cross section, which provides a semiclassical approximation  of the scattering cross section of the scalar wave in the KN black-bounce spacetime, is introduced  as \cite{Matzner:1985rjn}
	\begin{align}\label{eq:glory}
		\frac{\dd \sigma}{\dd \Omega}\simeq 2\pi\omega v b^2_g\Big|\frac{\dd b}{\dd \theta}\Big|_{\theta\simeq \pi}[J_{0}(\omega  v b_g\sin\theta)]^2,
	\end{align}
	where $b_g$ is called the glory impact parameter, and its value is defined as 
	\begin{align} \label{eq:bgdef} 
		b_g=b(\pi), 
	\end{align} 
	i.e., the impact parameter at which $\Delta\theta=\pi$ or when the signal turns back to the incoming direction. Here, $J_{0}$ is the 0th-order Bessel function of the first kind, whose appearance is essentially determined by the asymptotic form of the scattered wave, just as in elementary quantum mechanical scattering  \cite{Matzner:1985rjn,Newton:2013}.  Obviously, the above equation only considers the case $\Delta\theta=\pi$. This is because the contribution of the $N=0$ (one u-turn) case to glory scattering is the largest. 
	
	There are a few observations that we can make about the glory scattering cross section from Eq. \eqref{eq:glory}. We see that due to the argument of the function $J_0$, the glory scattering cross section will illustrate an oscillating feature as $\theta$ increases, with the width of each oscillation determined by the factor $\omega v b_g$ in the argument of the function. Therefore, the parameters $(a,\,Q,\,k,\,q)$ of the spacetime and wave will affect some features of this oscillation, as seen in Fig. \ref{fig:sca1}, through the factor $b_g$. Clearly, the larger the $b_g$, the faster the $J_0$ function will oscillate as $\theta$ increases. The factors in front of the $J_0$ function in Eq. \eqref{eq:glory} provide a smooth baseline for the $J_0$ oscillation. These observations will help us to better understand the features shown in Figs. \ref{fig:sca2} and \ref{fig:sca1}, where the effect of the spacetime and wave parameters on the cross section is analyzed. 
	
	\section{Partial wave approach} \label{sec:partial wave}
	In this section, we study the absorption and scattering of charged massive scalar fields in the KN black-bounce spacetime. We consider the perturbation of such a wave $\Psi_{\omega}$ with frequency $\omega$, which satisfies the following Klein-Gordon equation
	\begin{align}	\label{eq:kge}
		\left[\left(\nabla_{\alpha} - i q A_{\alpha}\right)\left(\nabla^{\alpha} - i q A^{\alpha}\right)-\mu^2\right]\Psi_{\omega} =0.
	\end{align}
	
	To facilitate the cross section computation,  we use the following ansatz, which allows the separation of variables in the Boyer-Lindquist coordinates
	\begin{align}	\label{eq:ansatz}
		\Psi_{\omega }(t,r,\theta,\phi)= \sum_{l=0}^{\infty}\sum_{m=-l}^{l}\frac{F_{\omega l m}(r)S_{\omega l m}(\theta)}{\sqrt{r^2+ k^2+a^2}} \expe^ {i m \phi- i\omega t},
	\end{align}
	where $l$ and $m$ are the angular quantum number and the azimuthal number, respectively. After substituting into Eq. \eqref{eq:kge} and separating of the variables, we find that the spheroidal harmonics $S_{\omega l m}(\theta)$ satisfy the following equation 
	\begin{align}\label{eq:spheroidal}
		&\left(\frac{\dd ^2}{\dd \theta^2}+\cot\theta\frac{\dd }{\dd \theta}\right)S_{\omega lm}\nonumber \\ 
		&+\left[\lambda_{lm}+a^2\left(\omega^2-\mu^2\right)\cos^2\theta-\frac{m^2}{\sin^2\theta}\right]S_{\omega lm}=0,
	\end{align}
	where $\lambda_{lm}$ represents the angular eigenvalue. The radial wave function $F_{\omega l m}(r) $ is subject to the following ordinary differential equation
	\begin{align}\label{eq:radial}
		\left(\frac{\dd ^2}{\dd  r_*^2}+V_{\omega lm}\right)F_{\omega lm}(r_*)=0,
	\end{align}
	where $r_*$ is  the tortoise coordinate linked with $r$ through the relation 
	\begin{align}\label{eq:tortoisel}
		r_{*}\equiv\int \dd r \left(\frac{r^2+ k^2+a^2}{\Delta}\right),
	\end{align}
	and $V_{\omega lm}$ is the potential given by 
	\begin{widetext}	\begin{align}\label{eq:veff}
			V_{\omega lm}(r)=&\frac{H^2+ \left[2ma\omega-\mu^2 \left(r^2+ k^2+a^2\right) - \lambda_{l m}-a^2\left(\omega^2-\mu^2\right)\right]\Delta}{\left(r^2+ k^2+a^2\right)^2}-\frac{\left[\Delta+2\sqrt{r^2+ k^2}\left(\sqrt{r^2+ k^2}-M\right)\right]\Delta}{\left(r^2+ k^2+a^2\right)^3} \nonumber \\  
			&+ \frac{3\left(r^2+k^2\right)\Delta^2}{\left(r^2+k^2+a^2\right)^4} - \frac{k^2\left[-4 M\left(r^2+k^2\right)+\sqrt{r^2+ k^2}\left(3Q^2+r^2+ k^2\right)+a^2\left(2M+\sqrt{r^2+ k^2}\right)\right]\Delta}{\sqrt{r^2+ k^2}\left(r^2+k^2+a^2\right)^4},
		\end{align}
	\end{widetext}
	with $H\equiv\left(r^2+k^2+a^2\right)\omega - am - q Q\sqrt{r^2+ k^2}$. We can check that when $r \rightarrow r_{h}$ (i.e. $r_*\rightarrow -\infty$), $V_{\omega lm}(r) \rightarrow \omega_{h} $ and when $r \rightarrow \infty$ (i.e. $r_*\rightarrow \infty$), $V_{\omega lm}(r) \rightarrow \omega_\infty $, where 
	\begin{align}  
		&\omega_{h} =\omega - \frac{a m+q Q \sqrt{r^2_h+ k^2}}{r_h^2+ k^2+a^2}\equiv \omega- \omega_c, \label{eq:omegacdef}  \\  
		&\omega_\infty \equiv \sqrt{\omega^2-\mu^2}. \label{eq:omegainfdef}
	\end{align}
	In the last step of Eq. \eqref{eq:omegacdef}, we defined a cutoff frequency
	\begin{align}
		\omega_c=\frac{a m+q Q \sqrt{r^2_h+ k^2}}{r_h^2+ k^2+a^2}.
	\end{align}
	Note that $\omega_c$ is independent of the partial wave index $l$. 
	
\begin{figure*}[htp!]
		\includegraphics[width=0.48 \textwidth]{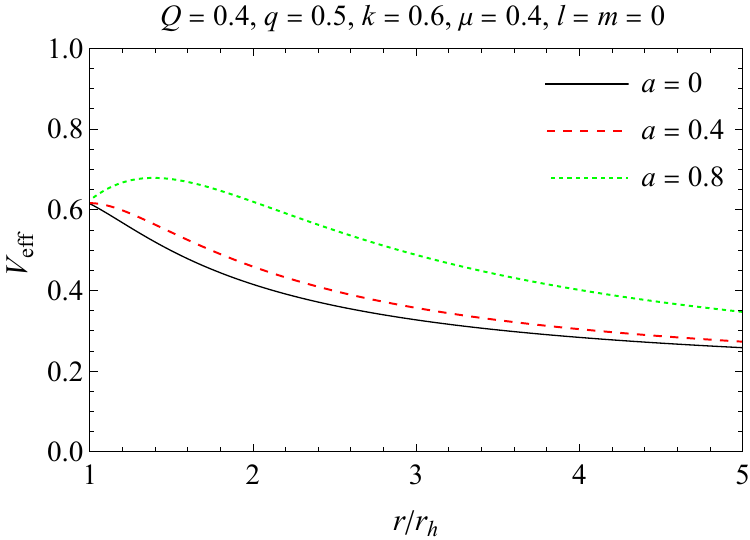}\hfill\includegraphics[width=0.48 \textwidth]{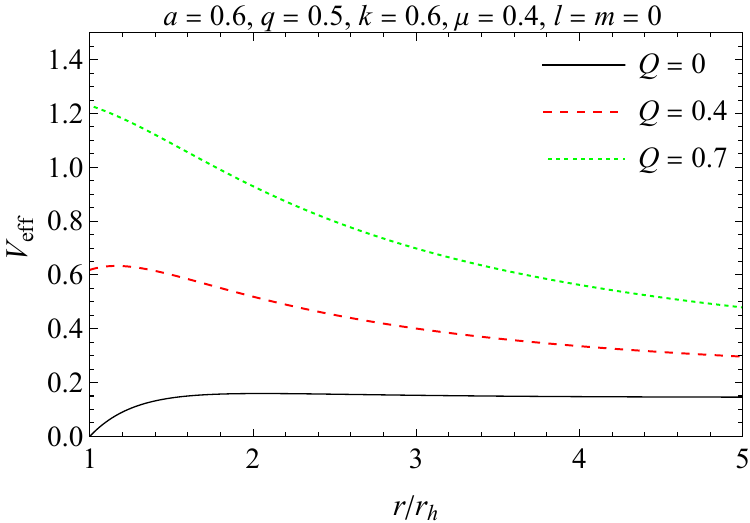}
		\includegraphics[width=0.48 \textwidth]{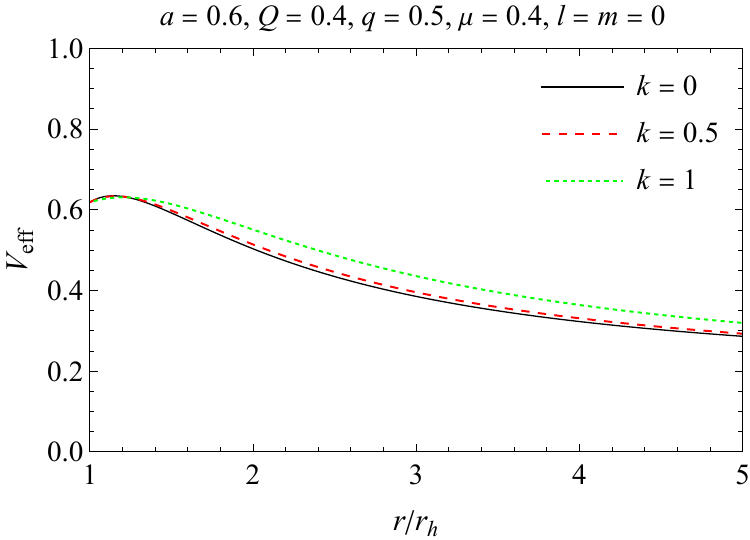}\hfill\includegraphics[width=0.48 \textwidth]{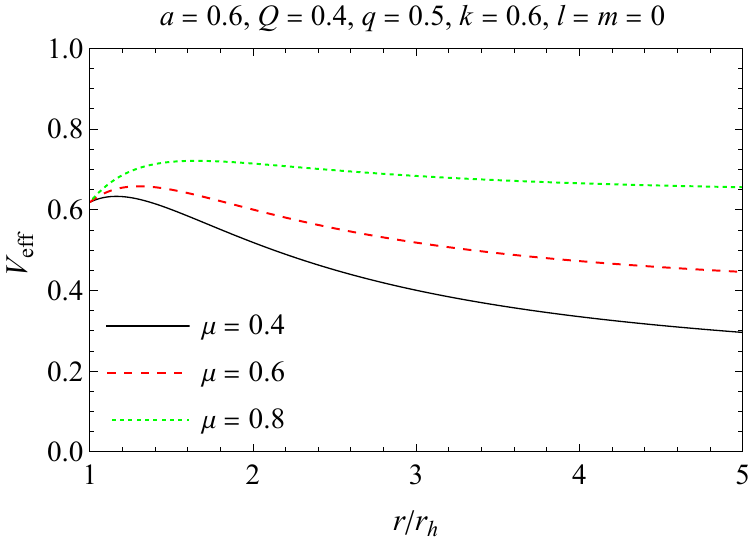}
		\includegraphics[width=0.48 \textwidth]{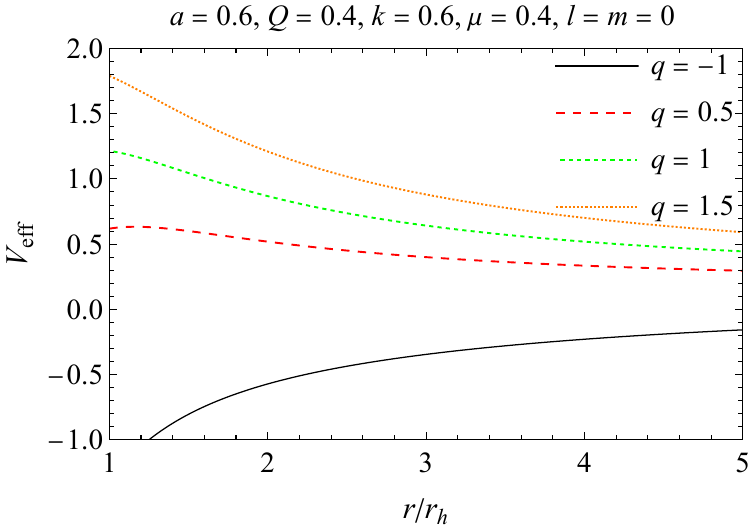}
		\caption{The effective potential $V_{\text{eff}}$ of the KN black-bounce spacetime for different values of $a$ (top left), $Q$ (top right), $k$ (middle left), $\mu$ (middle right), and $q$ (bottom).} \label{fig:V}
	\end{figure*}

	In order to better understand the behavior of the cross sections, we show the effective potential, $V_{\text{eff}} = -V_{\omega l m} + \omega^{2}$, as a function of the parameters $a$, $Q$, $k$, $\mu$, and $q$ in Fig. \ref{fig:V}. We find that the peak of the effective potential increases with $a$, $Q$, $\mu$, and $q$, but decreases with $k$. Moreover, we observe that the parameter $k$ has a smaller effect on the effective potential than the other parameters.

	Since we are interested in the absorption and scattering of the charged massive scalar wave under the effective potential, we consider the following asymptotic form of the wave  \cite{Futterman:1988}	\begin{align}\label{eq:solution}
		F_{\omega lm}(r) \approx
		\left\{
		\begin{array}{ll}
			\expe^ {-i\omega_\infty r_*} + \mathcal{R}_{\omega l m} \expe^ {i\omega_\infty r_*}, \quad &\mbox{for $r_*\rightarrow +\infty$},\\
			\mathcal{T}_{\omega l m}   \expe^ {-i \omega_{h} r_*}, &\mbox{for $r_*\rightarrow -\infty~$},
		\end{array}
		\right.
	\end{align}
	where  $\mathcal{R}_{\omega l m}$ and $\mathcal{T}_{\omega l m}$ are the reflection and transmission coefficients, respectively, and they satisfy the relation
	\begin{align}\label{eq:rtf}
		|\mathcal{R}_{\omega lm}|^2 + \frac{\omega_{h}}{\omega_\infty}|\mathcal{T}_{\omega lm}|^2 = 1.
	\end{align}
	From definition \eqref{eq:omegacdef} and Eq. \eqref{eq:solution}, it is evident that when $\omega < \omega_c$, the quantity $\omega_h$ undergoes a sign reversal, causing the test scalar wave to propagate in the opposite direction (outward) at the event horizon. This phenomenon leads to wave amplification, a process known as superradiance, which will be thoroughly examined in Sec. \ref{sec:superradiance}.  Furthermore, the condition $\omega>\mu$ obtainable from Eq. \eqref{eq:omegainfdef} for the scattering waves to reach infinity needs to be satisfied in order to study the scattering properties.
	
	Since we focus on a plane wave propagating along the $\hat{z}^+$ direction, i.e., on-axis incidence with index $m=0$,  the total absorption cross section $\sigma_{\text{abs}}$ for such a wave is given by \cite{Futterman:1988} 
	\begin{align} \label{eq:tacs}
		\sigma_{\text{abs}} =\sum_{l=0}^{\infty}\sigma_{l,0},
	\end{align}
	where $\sigma_{l,0}$ refers to the partial absorption cross section
	\begin{align} \label{eq:pacs}
		\sigma_{l,0}= \frac{4\pi^2}{\omega_\infty^2}\left|S_{\omega l0}(0) \right|^2 \left(1-\left|\mathcal{R}_{\omega l0} \right|^2 \right).
	\end{align}
	From Eqs. \eqref{eq:omegacdef} and \eqref{eq:rtf} and the fact that $\omega_c$ is $l$-independent, we see that when $\omega_h<0$ or, equivalently, $\omega<\omega_c$, all $(1-|\mathcal{R}_{\omega l 0}|^2)$ factors in Eq. \eqref{eq:pacs} and consequently all partial absorption cross sections $\sigma_{l,0}$ with different $l$, as well as the total absorption cross section $\sigma_{\text{abs}}$ in Eq. \eqref{eq:tacs}, simultaneously become negative. This is one of the most prominent features or characteristics demonstrating that superradiance has occurred in this scattering process.

	Now, for the differential scattering cross section of this wave, it is given by  \cite{Futterman:1988}
	\begin{align} \label{eq:dscs}
		\frac{\dd \sigma}{\dd \Omega} =\left|f(\theta)\right|^2,
	\end{align}
	where the scattering amplitude $f(\theta)$ takes the form
	\begin{align}\label{eq:sa}
		f(\theta) =\frac{2\pi}{i\omega_\infty}\sum_{l=0}^{\infty}S_{\omega l0}(0)S_{\omega l0}(\theta) \left[(-1)^{l+1}\mathcal{R}_{\omega l0}-1\right].
	\end{align}
	Here the $S_{\omega l 0}$ was given in Eq. \eqref{eq:spheroidal}.
	Using expressions \eqref{eq:tacs} and  \eqref{eq:dscs}, in the next section, we will present the numerical results for the absorption and differential scattering cross sections and their dependence on various parameters, paying special attention to the superradiance situation in Sec. \ref{sec:superradiance}. 
	
	\section{numerical results and analysis}\label{sec:numerical results}
	
	In this section, we discuss the absorption and scattering cross sections obtained by the partial wave method and compare them with analytical results (geodesic scattering \eqref{eq:cdscs} and glory scattering \eqref{eq:glory}). It is clear from Eqs. \eqref{eq:pacs} and \eqref{eq:sa} that we only need to calculate the reflection coefficient $\mathcal{R}_{\omega l0}$ and the spheroidal harmonics $S_{\omega l 0}$. For $\mathcal{R}_{\omega l0}$, we use the fourth-order Runge-Kutta method to solve the radial wave equation \eqref{eq:radial}, a second-order differential equation, with the calculation process detailed in Ref. \cite{Dolan:2009zza}. To solve the angular equation \eqref{eq:spheroidal} and obtain the eigenvalues $\lambda_{l0}$, we use the spectral decomposition method \cite{Hughes:1999bq}. Note that the scattering amplitude \eqref{eq:sa} is poorly convergent. Thus, we adopt a series reduction technique to address this issue \cite{Leite:2019eis,Leite:2019zqo}.
	
	\subsection {Absorption cross section}
	
	\begin{figure*}[htp!]
		\includegraphics[width=0.48 \textwidth]{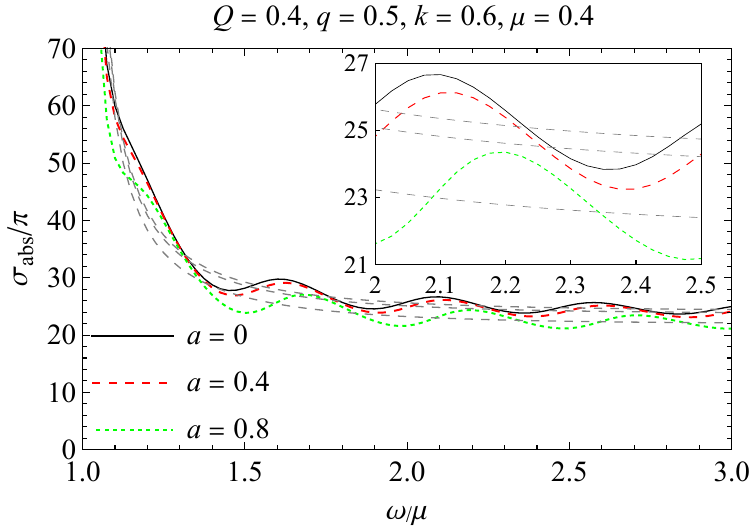}\hfill\includegraphics[width=0.48 \textwidth]{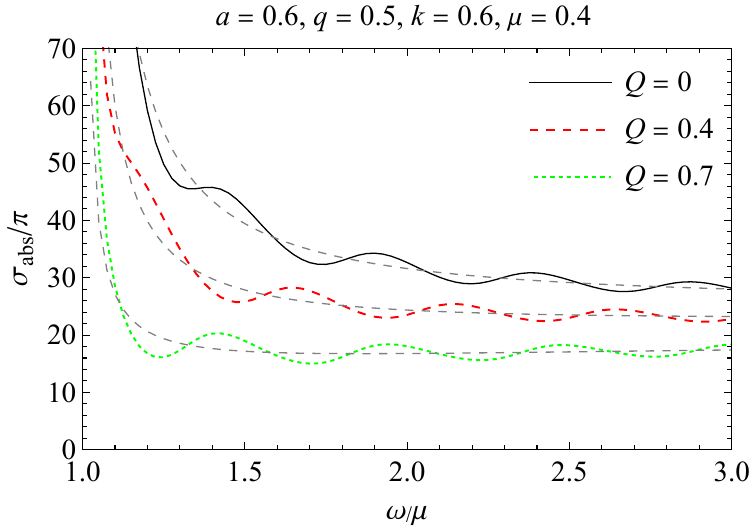}
		\includegraphics[width=0.48 \textwidth]{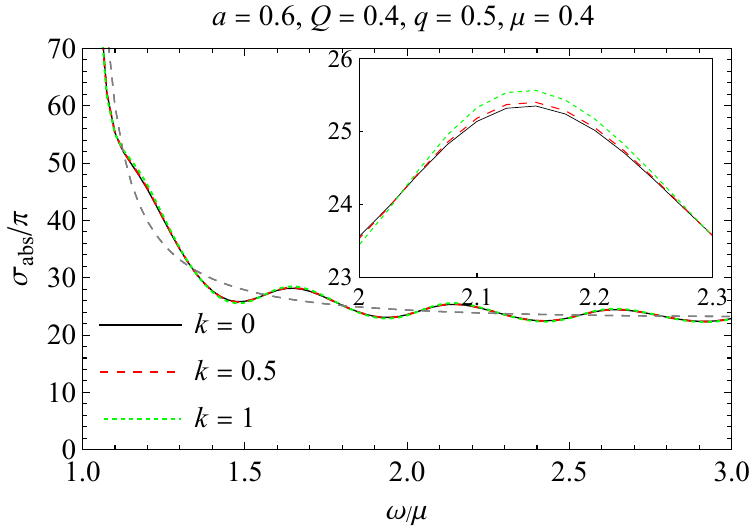}\hfill\includegraphics[width=0.48 \textwidth]{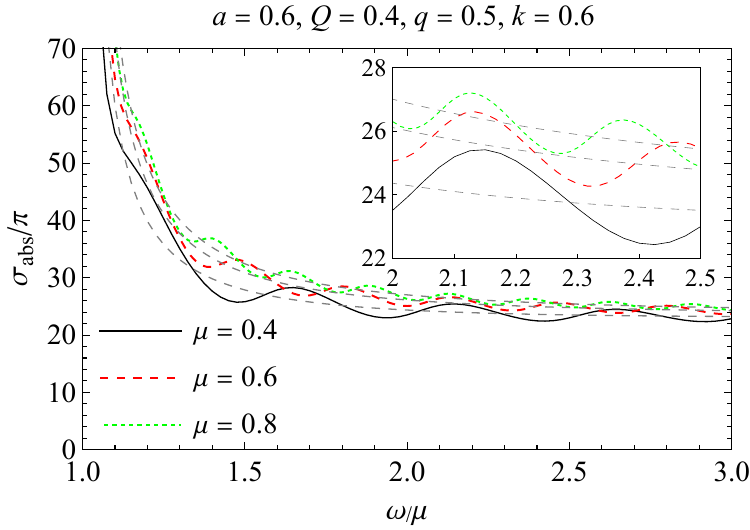}
		\includegraphics[width=0.48 \textwidth]{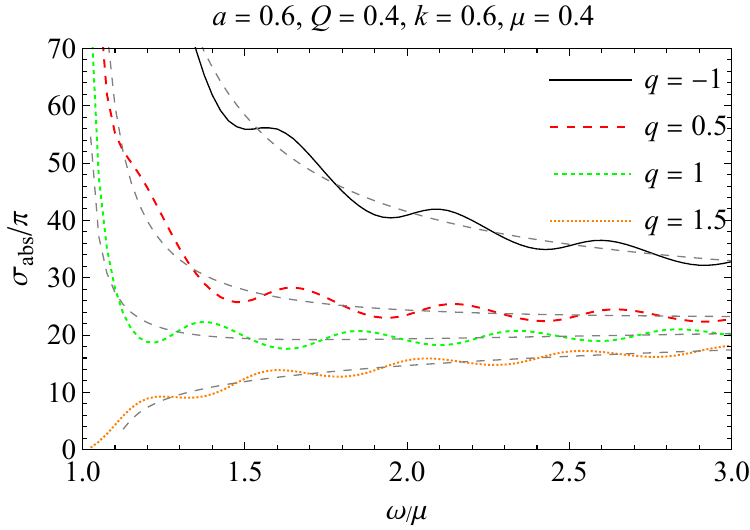}
		\caption{The total absorption cross section of the KN black-bounce spacetime for different values of $a$ (top left), $Q$ (top right), $k$ (middle left), $\mu$ (middle right) and $q$ (bottom). The gray dashed line represents the value of the geometric cross section. The inserts are a localized total absorption cross section and geometric cross section  near $\omega/\mu=2.2$.} \label{fig:abs1}
	\end{figure*}
	
	Let us start by presenting the total absorption cross section of the charged massive scalar field in Fig. \ref{fig:abs1}. We note from all plots that the numerical results (all curves except the gray dashed lines) obtained by the partial wave method as in Eq. \eqref{eq:tacs}, regularly oscillate around the geometric cross sections (the gray dashed lines) obtained using Eq. \eqref{eq:gcs} in the high-frequency regions.   This is because waves incident along the axis of rotation are not subject to the frame-dragging effect \cite{Macedo:2013afa}. Therefore, this feature is similar to the behavior of the total absorption cross section of a static BH  \cite{Benone:2015bst}.

	Regarding the effects of the spacetime parameters on the absorption cross section, we first observe from the top panels in Fig. \ref{fig:abs1} that increasing the values of spacetime spin $a$ and charge $Q$ causes the total absorption cross section to decrease.   The reason for this is that an increase in spacetime spin $a$ and charge $Q$ strengthens the effective potential, as seen from Fig. \ref{fig:V} top panels, and thus suppresses wave transmission.  Besides, the effect of $Q$ here is similar to the effect of $Q$ in the RN BH and charged Horndeski BH cases \cite{Pang:2018jpm,Li:2024xyu}.
	Then we note from the middle left plot that enhancing the regularization parameter $k$  has a different effect.  It does not modify the geometric cross section (see Fig. \ref{fig:rc-bc}), but leads to a slight enhancement in the amplitude of the total absorption cross section oscillation, analogous to the phenomenon observed in the black-bounce spacetime \cite{Lima:2020auu}.   This is because the parameter $k$ reduces the peak value of the potential, and its effect on the effective potential is weaker than that of the other parameters, as shown in the middle left frame of Fig. \ref{fig:V}. Moreover, as the angular momentum $l$ increases, the influence of $k$ on $\sigma_{l0}$ becomes less pronounced (see Fig. 9 of Ref.~\cite{Lima:2020auu}), indicating that $k$ predominantly affects $\sigma_{00}$. However, the introduction of field mass causes $\sigma_{00}$ to diverge \cite{Benone:2014qaa}, making the most influential part ``invisible.'' Consequently, the regularization parameter $k$ modifies the absorption cross section much more weakly compared to the other parameters.

	The effect of the remaining parameters on the absorption cross section is now going to be analyzed. The middle right and bottom plots show that the absorption cross section increases with the increase of field mass $\mu$ or the decrease of field charge $q~(>0)$, indicating that a heavier or less repulsive field is easier to absorb.  	Generally, the higher the potential barrier, the fewer waves are transmitted. However, we find that as the field mass increases, both the height of the potential barrier and the absorption cross section increase, which is contrary to our initial expectations. This behavior stems from the competition between two factors: the denominator $\omega_\infty^2$ and the numerator $\left|S_{\omega l0}(0)\right|^2\left(1-\left|\mathcal{R}_{\omega l0}\right|^2\right)$. While both factors decrease with increasing field mass $\mu$, the $\omega_\infty^2$ term decays more rapidly \cite{Jung:2004yh}.  Moreover, note from the $\omega_\infty^2$ factor in Eq.  \eqref{eq:pacs} that as $\omega\to\mu^+$, the absorption cross section tends to infinity for neutral massive scalar waves because $|\mathcal{R}_{\omega l0}|<1$ in this case. This means that a neutral field with energy very close to its mass, and therefore little asymptotic kinetic energy, will eventually always wind into the horizon. When $qQ<0$, one can also easily check using $b_c$ in Eq. \eqref{eq:gcs} that the absorption cross section will be enhanced compared to the neutral field case and will further increase as $|qQ|$ increases for all $\omega$, including when $\omega\to\mu^+$. This is intuitively expected because the central charge causes extra attraction on top of the gravitational attraction to the field in such cases. In contrast, however, if the electromagnetic force ($q Q > 0$)  is present and large enough, the factor $(1-|\mathcal{R}_{\omega l0}|^2)$ in Eq. \eqref{eq:pacs} can also be very small (refer to Fig. \ref{fig:sur_am_aqk}) and consequently renders the absorption cross section less divergent in the same $\omega\to\mu^+$ limit, as seen from the bottom plot of Fig. \ref{fig:abs1}.

	\subsection {Scattering cross section}
	
	\begin{figure}[htp!]
		\centering
		\includegraphics[width=0.48\textwidth]{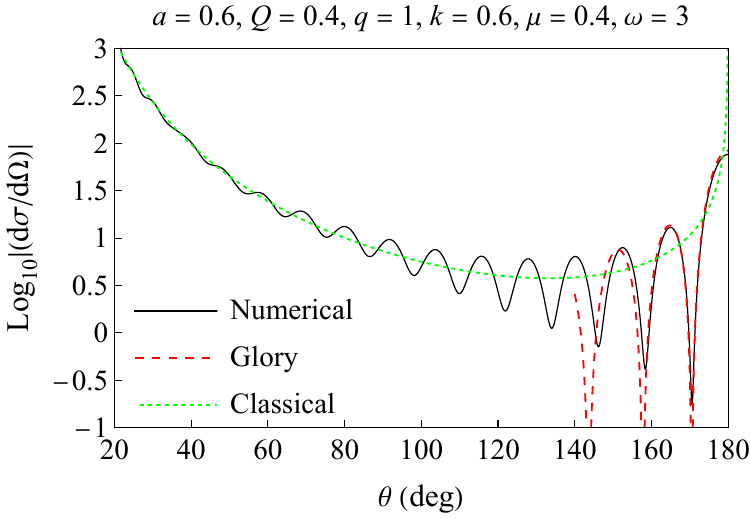}
		\caption{Comparison of the numerical result obtained by the partial wave method with geodesic scattering and glory scattering.}	\label{fig:sca2}
	\end{figure}

	In Fig. \ref{fig:sca2}, we present the differential scattering cross section computed numerically using the partial wave method formula \eqref{eq:dscs},  along with the classical and semiclassical scattering cross sections given by Eqs. \eqref{eq:cdscs} and \eqref{eq:glory}, respectively. We note that both the numerical and classical results show divergent behavior in the forward direction, agreeing with both the gravitational scattering of the uncharged scalar field  \cite{Leite:2019eis} and the classical scattering of charged particles by the  Coulomb potential \cite{Hamilton:2010}. We also observe that the more the scattering angle approaches the backward direction, the better the glory scattering cross section approximates the numerical results, indicating that glory scattering is reliable for analyzing the dependence of the scattering cross section on other parameters when the scattering angle is large. 
	
	\begin{figure*}[htp!]
		\includegraphics[width=0.48 \textwidth]{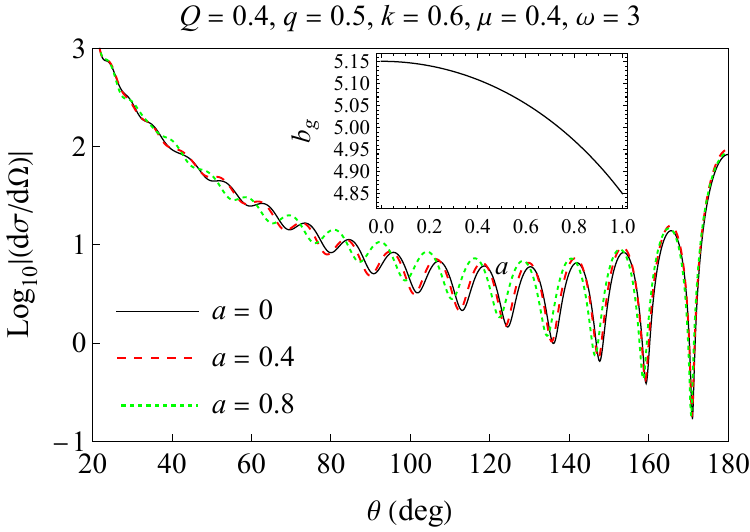}\hfill\includegraphics[width=0.48 \textwidth]{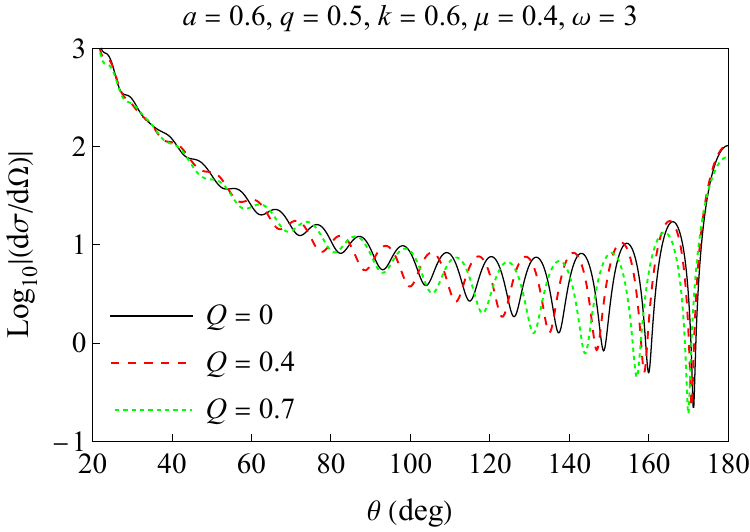}
		
		\includegraphics[width=0.48 \textwidth]{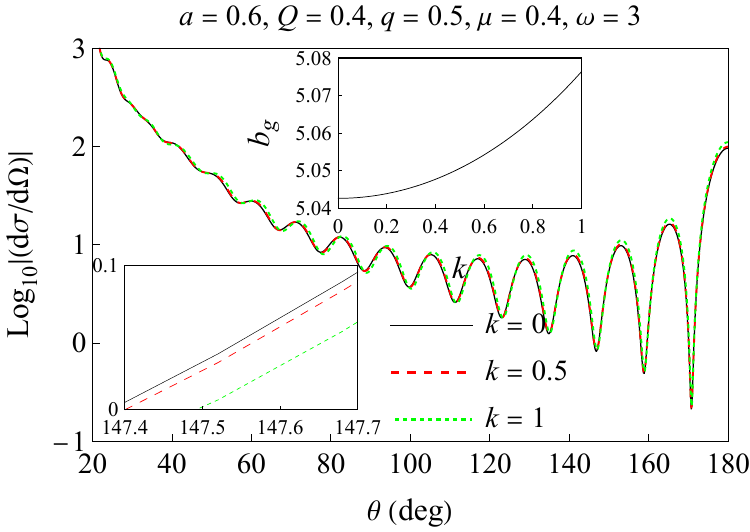}\hfill\includegraphics[width=0.48 \textwidth]{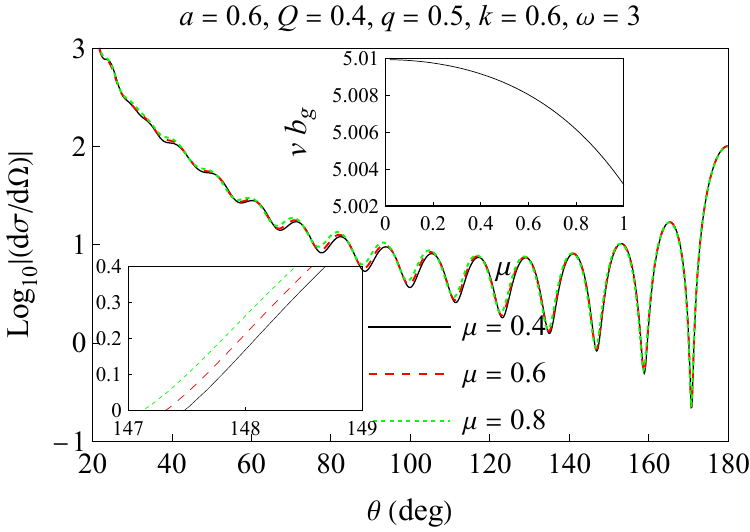}
		
		\includegraphics[width=0.48 \textwidth]{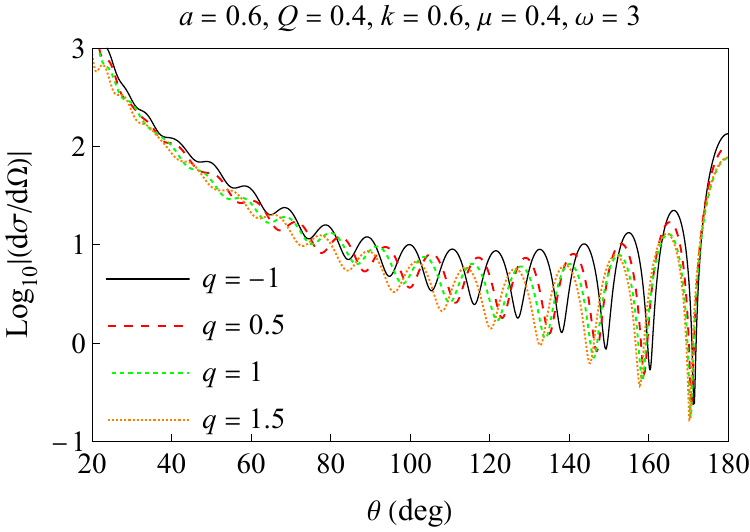}\hfill\includegraphics[width=0.48 \textwidth]{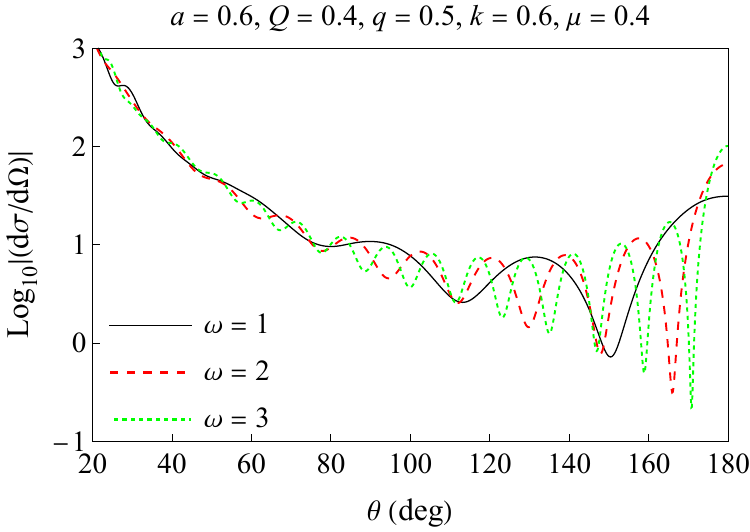}
		\caption{The differential scattering cross section of the KN black-bounce spacetime for different values of $a$ (top left), $Q$ (top right), $k$ (middle left), $\mu$ (middle right), $q$ (bottom left) and $\omega$ (bottom right).  The insets at the bottom left corner of the middle plots are used to better see the effect of the parameters $k$ and $\mu$ on the differential scattering cross section. }	\label{fig:sca1}
	\end{figure*}
	
	In Fig. \ref{fig:sca1}, we show the differential scattering cross section as a function of $\theta$ for different values of the parameters $(a,\, Q,\,k,\,\mu,\,q)$ and frequency $\omega$. In general, we can see from all the plots the oscillations of the scattering cross section at the intermediate angles as well as a scattering peak at $\theta=\pi$, which are features similar to the case of neutral scalar wave scattering.
	
	For the effect of individual parameters on the scattering cross section, we first observe from the top plots, the middle right and bottom left plots of Fig. \ref{fig:sca1} that as $a~(>0),\, Q~(>0)$~(when $q>0),\,q~(>0)$~(when $Q>0)$ or $\mu$ increase, the interference fringes of the differential scattering cross section become wider.  The middle left plot, on the other hand, illustrates that the introduction of the regularization parameter $k$ narrows the width of the interference fringes, although this effect is weaker compared to that of $a,\, Q$ and $q$ for the chosen ranges of parameters.  As pointed out in Sec. \ref{subsec:Glory}, these features of the cross section can be well understood from the effect of the corresponding parameters on the various arguments/factors of Eq. \eqref{eq:glory}, especially in the $\theta\to\pi$ limit. It was previously known that to maintain the deflection angle of the (charged) rays at $\pi$, the impact parameter $b_g$ will have to decrease if the spacetime charge $Q~(>0)$ increases  \cite{Pang:2018jpm,Zhou:2022dze}, or the field charge $q~(>0)$ increases \cite{Zhou:2022dze}, which explains the effect of $Q$ and $q$ mentioned above. For the spacetime spin $a$ and parameter $k$, few studies have examined their effect on $b_g$ when the rays propagate along the $\hat{z}^+$ direction. The numerical study in this work shows that indeed the increase of $a$ or the decrease of $k$ will reduce $b_g$, as illustrated in the insert at the upper right  corner of the top left and middle left plots of Fig. \ref{fig:sca1}. Moreover, a close inspection of the middle left plot shows that the average value of the differential scattering cross section, which roughly matches the classical scattering cross section, increases slightly as $k$ increases (most apparent from the peak heights of the oscillations). This is in agreement with the effect of the regularization parameter in other black-bounce spacetimes \cite{LimaJunior:2022zvu}. When the wave mass $\mu$ increases but the wave frequency $\omega$ is fixed, the wave will have a smaller asymptotic velocity $v$. Therefore, a simple intuition suggests that $b_g$ will have to increase in order for the deflection to still reach $\pi$. The product of $v b_g$, as in the argument of the $J_0$ function in Eq. \eqref{eq:gcs}, turns out to decrease (see the insert at the upper right  corner  of the middle right plot of Fig. \ref{fig:sca1}) as $\mu$ increases, thereby forcing the oscillation fringe width to become larger.
	
	Besides the above effects, there is another feature of interest in the bottom left plot that the amplitude of the differential scattering cross section at medium-large scattering angles decreases as $qQ$ increases, indicating that a more repulsive interaction causes the spacetime to bend more signal away from but less signal back to the source direction. 
	With regard to the influence of frequency, we observe that as the incident frequency increases, the width of the interference fringes decreases while the amplitude of the oscillations increases.  The width of the oscillations is inversely proportional to the coefficients $\omega v b_{g}$, as shown in Sec. \ref{subsec:Glory}.  That is to say, the higher the incident wave frequency, the narrower the width of the interference fringes.

	As can be observed from Fig. \ref{fig:sca1}, the parameter $q$ exhibits the most pronounced influence on the fringe width of the differential scattering cross section and the average scattering flux. This is fundamentally attributed to the fact that the potential barrier height is most sensitive to variations in the parameter $q$. While physical intuition suggests that a higher potential barrier would lead to increased wave reflection and thus enhanced scattering flux, the expression for the differential scattering cross section reveals a more complex, nonlinear relationship—the differential scattering cross section is not a monotonic function of the reflection coefficient but rather exhibits quadratic characteristics. Furthermore, at intermediate to large scattering angles, higher-angular-momentum partial waves undergo spiral scattering around the BH, because lower-angular-momentum partial waves are predominantly absorbed by the BH \cite{Anninos:1992ih}. This helical mechanism further modulates the final scattering flux.

	\section{Superradiance} \label{sec:superradiance}
	
	Penrose \cite{Penrose:1971uk} presented the extraction of rotational energy from a rotating BH, known as the Penrose process. Later, Misner \cite{Misner} demonstrated for the first time that when $\omega < m a/\left(r_h^2+a^2\right)$, the amplitude of the scattered wave is amplified, a phenomenon called superradiance, thus extracting energy from the rotating BH. Building on Hawking's theorem, Bekenstein \cite{Bekenstein:1973mi} revealed the existence of the Misner process, in which a charged field is amplified when scattered by a charged BH, thereby extracting charge and Coulomb energy from the BH. Gibbons considered the Penrose process and superradiance to be complementary aspects of ``wave-particle duality," which can be explained in a unified way by quantum field theory \cite{Gibbons:1975kk}.

	\subsection{Superradiant effect} \label{subsec:superradiance}
	From the discussion in Sec. \ref{sec:partial wave}, it was clear that  superradiance will occur for an on-axis scattering when 
	\begin{align} 
		\omega< \omega_c= \frac{q Q \left[M+\sqrt{M^2-(a^2+Q^2)}\right]}{2M^2-Q^2+2M\sqrt{M^2-(a^2+Q^2)} }. 
		\label{eq:omegacsub}
	\end{align}
	Here we have substituted the $r_h$ in Eq. \eqref{eq:rh} into $\omega_c$ as defined in Eq. \eqref{eq:omegacdef}. A few comments about this $\omega_c$ and the superradiance are in order here. The first is that superradiance will not happen for a neutral scalar field or spacetime, or when the Lorentz force between them is attractive ($qQ\leq 0$). Second, regarding the effect of the various parameters $(M,\,q,\,Q,\,a,\,k)$ on $\omega_c$, we see from Eq. \eqref{eq:omegacsub} that $\omega_c$ does not depend on $k$, but is proportional to $q^1$. Therefore, this is indeed the same as the critical frequency for the KN spacetime for a charged scalar. Moreover, through a simple analysis of $\omega_c$, we can show that it will increase monotonically with respect to the increase of $|Q|$ or $|a|$,  while $M$ can be thought of as providing a base scale against which other quantities and $\omega_c$ can be compared.

	\begin{figure*}[htp!]
		\centering
		~~~~\includegraphics[width=0.295\textwidth]{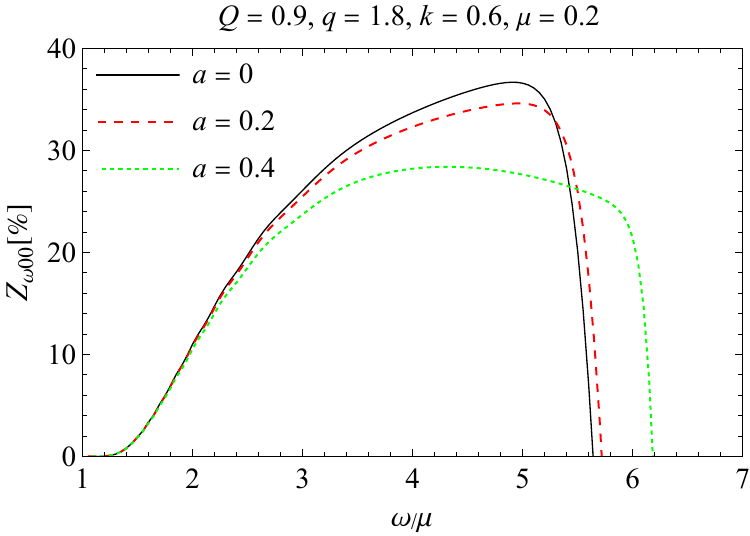}\hfill\includegraphics[width=0.295\textwidth]{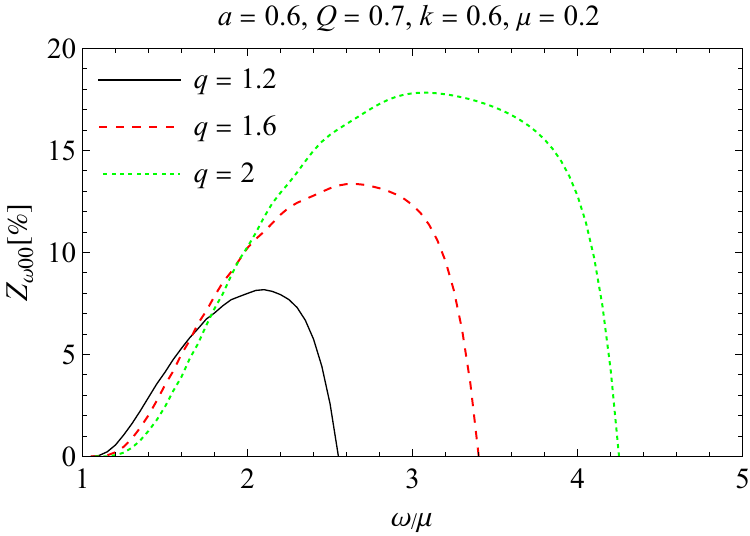}\hfill \includegraphics[width=0.295\textwidth]{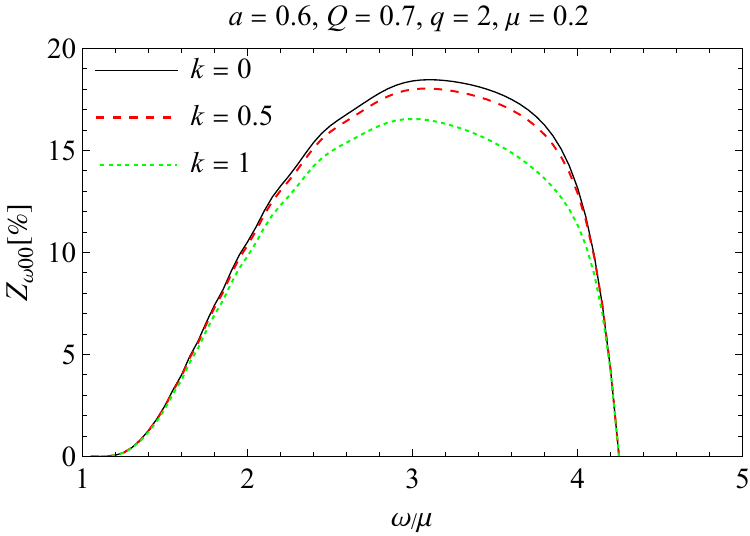} 
		
		\hspace{1mm}\includegraphics[width=0.309 \textwidth]{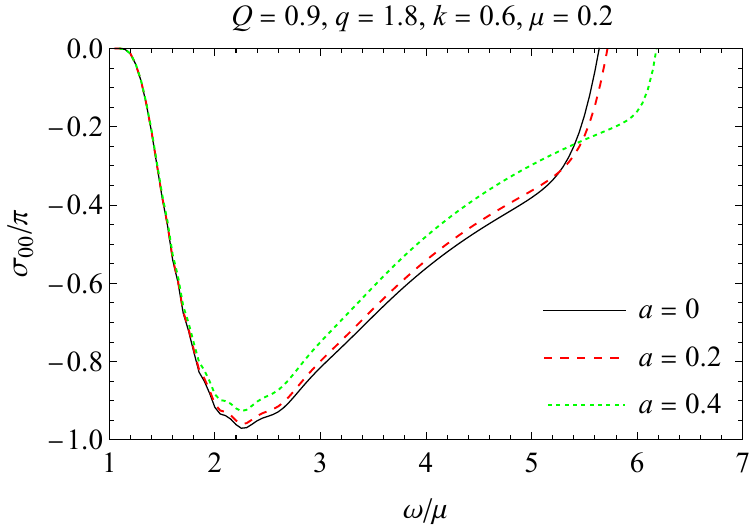}\hfill\includegraphics[width=0.308 \textwidth]{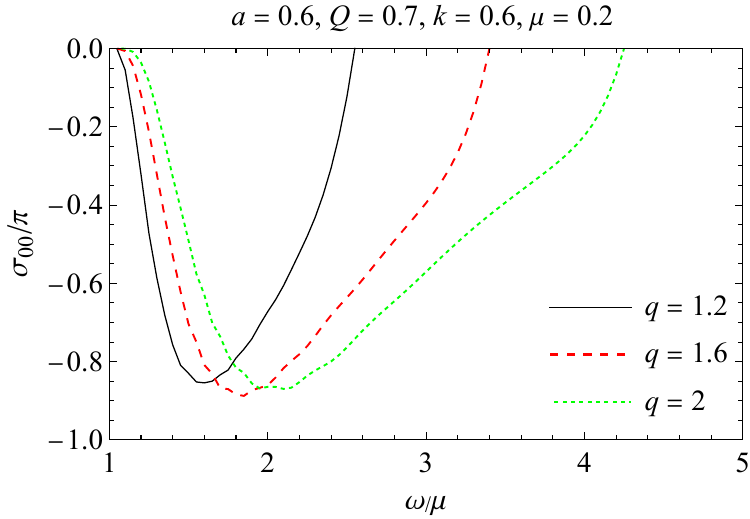}\hfill\includegraphics[width=0.308 \textwidth]{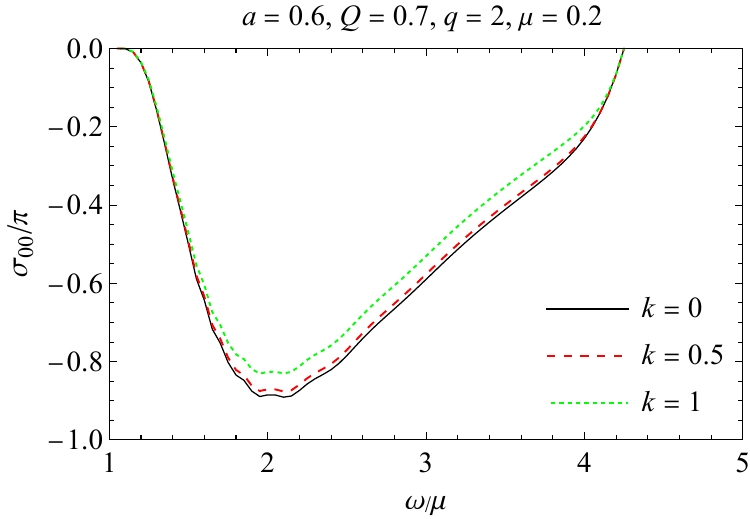}
		\caption{The amplification factor (top) and partial absorption cross section (bottom) of the KN black-bounce spacetime for $m=l=0$  for different values of  $a$ (left), $q$ (center) and $k$ (right).\label{fig:sur_am_aqk}}
	\end{figure*}

	When superradiance happens, we can measure its extent quantitatively using the absorption cross section $\sigma_{\text{abs}}$ and the amplification factor 
	\begin{align}
		Z_{\omega l 0}=|\mathcal{R}_{\omega l 0}|^2 -1=\frac{\dd E_{\text{out}}}{\dd E_{\text{in}}}-1,
	\end{align} 
	which also equals the energy amplification of the impinging planar wave at infinity. In Fig. \ref{fig:sur_am_aqk}, we present the amplification factor $Z_{\omega 00}$ (top plots) and the corresponding partial absorption cross section $\sigma_{00}$ (bottom plots) defined through Eq. \eqref{eq:pacs}, with respect to the change in frequency for different values of spin $a$ (left plots), charge $q$ (center plots) and regularization parameter $k$ (right plots). Note that similar to the scattering in other charged spacetimes \cite{dePaula:2024xnd}, the $l=0$ partial wave absorption cross section dominates the total one (roughly 95\%) under the parameter settings in the figure.  Therefore,  $\sigma_{00}$ can be roughly regarded as $\sigma_{\text{abs}}$ in the following analysis. 
	
	Firstly, we see that all plots in this figure show the existence of cutoff frequencies $\omega_c$ below which $Z_{\omega 0 0}$ (top plots) becomes positive and the corresponding $\sigma_{00}$ becomes negative (bottom plots). Moreover, we can numerically verify that the values of these $\omega_c$'s and their dependence on $a,\,q$ and $k$ match exactly the prediction of Eq. \eqref{eq:omegacsub}. In particular, from the right two plots, it is apparent that the $\omega_c$'s do not depend on the value of $k$.
	
	As $\omega$ changes, we observe from plots of each column that both $Z_{\omega 00}$ and $\sigma_{00}$ contain peaks at some frequencies between $\mu$ and $\omega_c$. Notably, the peaks of $\sigma_{00}$ occur at lower frequencies compared to those of $Z_{\omega 00}$. 
	The reason for this is that when we compute $\sigma_{00}$ from $Z_{\omega 00}$, an extra factor $\frac{4\pi^2}{\omega_\infty^2}\left|S_{\omega l0}(0) \right|^2$ as in Eq. \eqref{eq:pacs} has to be taken into account. This factor approaches infinity as $\omega\to\mu^+$, i.e., $\omega_\infty\to0$, and therefore effectively modifies the diminishing speed of $Z_{\omega 00}$ as $\omega$ approaches $\mu^+$ and moves the peak to lower frequencies. One further observes that the amplification at the peaks is around $\sim$8\% - $\sim$40\% depending on the values of $a,\,q$ and $k$. These amplification levels are significantly higher than those observed for scalar waves in the absence of Lorentz interactions (i.e., neutral scalar fields or Kerr spacetime) \cite{Teukolsky:1974yv,Brito:2015oca}.
	
	From the left column plots, it is evident that for on-axis incidence, the peak value of the amplification factor, as well as its value at fixed small $\omega$, decreases as the spin parameter $a$ of the spacetime increases. This observation stands in sharp contrast to the behavior of neutral scalar waves propagating along the equatorial plane in Kerr spacetime, where the amplification factor grows with increasing $a$ up to its extremal limit \cite{Liu:2024qso, Brito:2015oca}. Regarding the effect of charge $q$, the center column plots reveal that both the width of the frequency range and the peak value of the amplification increase with $q$. This suggests that the superradiance effect strengthens as the repulsive electrostatic force between the incoming wave and the spacetime becomes more pronounced, consistent with physical intuition.  However, it is noteworthy that at the low-frequency end, the order reversal of the amplification factor indicates that frequency dependence outweighs the Lorentz force effect in this regime. Consequently, while the peak values of $Z_{\omega 00}$ grow with Lorentz repulsion, the peak of $\sigma_{00}$ first increases and then decreases. This phenomenon demonstrates that the superradiant intensity cannot be simply characterized by the peak value of $\sigma_{00}$. As for the effect of the regularization parameter $k$, the plots in the right column demonstrate that the peak values of superradiance decrease as $k$ increases. To the best of our knowledge, this is the first study to investigate the influence of the black-bounce parameter $k$ on superradiance in the presence of electromagnetic interaction. This is consistent with the results for a neutral scalar wave scattered by the Kerr black-bounce spacetime \cite{Franzin:2022iai}.

	\begin{figure*}[htp!]
		\includegraphics[width=0.48 \textwidth]{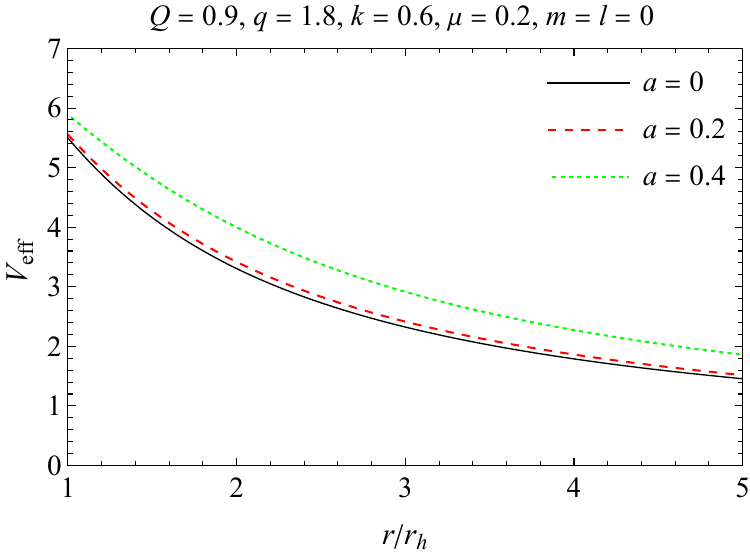}\hfill\includegraphics[width=0.48 \textwidth]{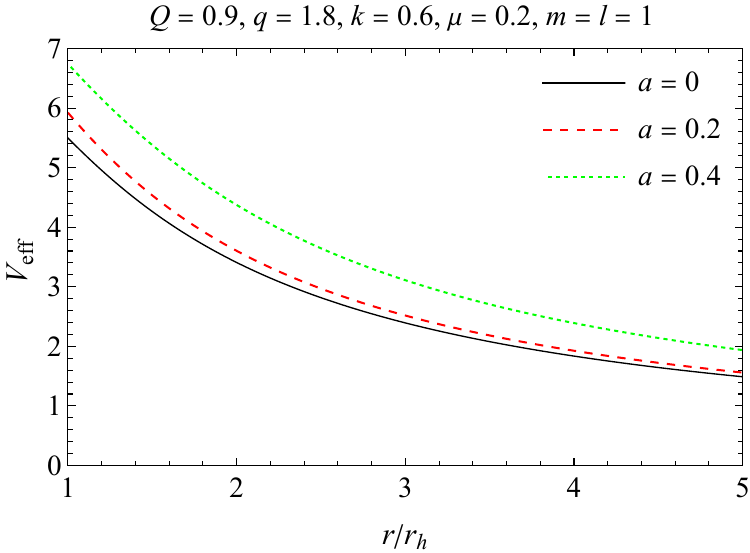}
		\includegraphics[width=0.48 \textwidth]{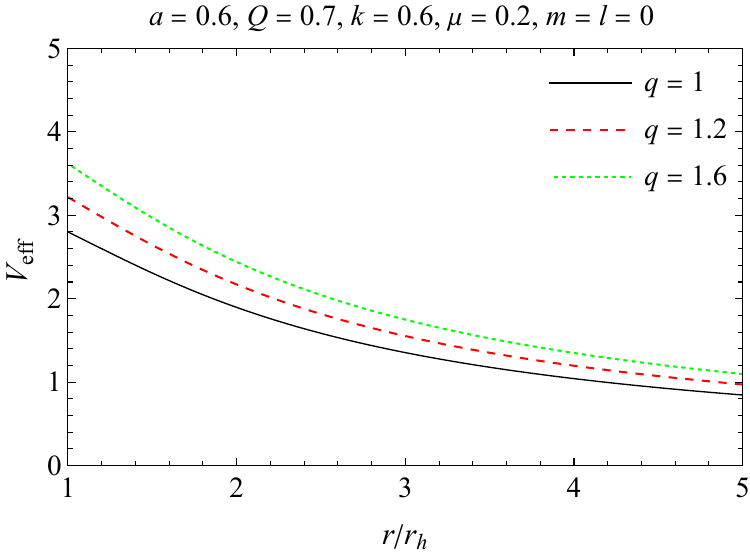}\hfill\includegraphics[width=0.48 \textwidth]{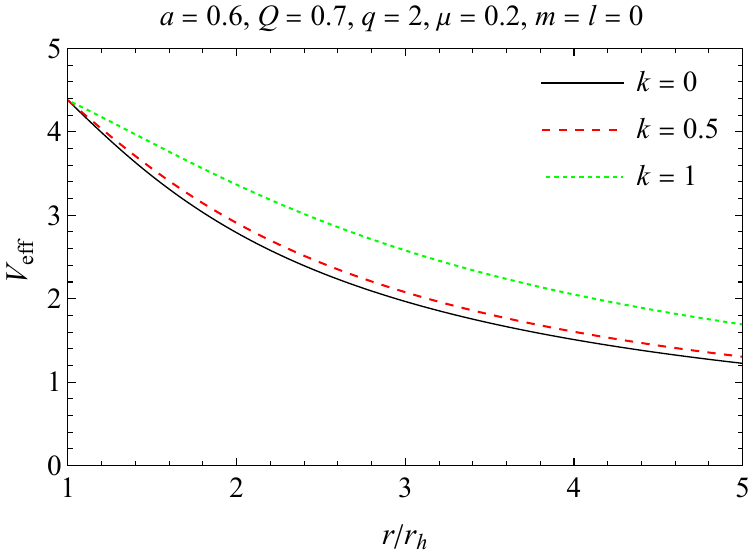}
		\caption{The effective potential  $V_{\text{eff}}$ of the KN black-bounce spacetime for different values of $a$ (top), $q$ (bottom left) and $k$ (bottom right) .} \label{fig:V1}
	\end{figure*}
	
	Since absorption and scattering cross sections are fundamentally governed by the potential barriers surrounding BHs, superradiance might be similarly regulated by these barriers. Zhang \textit{et al.}~\cite{Zhang:2020sjh} have explained parameter-dependent effects on superradiance by analyzing variations in the effective potential. Based on this, we also attempt to understand changes in superradiance in terms of the potential. In Fig.~\ref{fig:V1}, we present the effective potential $V_{\text{eff}}$, where we vary the parameters $a$, $q$, and $k$ for $l=m=0$ when superradiance occurs. To compare the effect of the spin parameter on superradiance in the equatorial case, we also show the effective potential for $l=m=1$ under variations of $a$. We find that the effective potential increases with larger values of the parameters $a$, $q$, and $k$. Therefore, we cannot explain the parameter dependence of superradiance by analyzing variations in the effective potential  as in Ref. \cite{Zhang:2020sjh}. 
	
	Some studies have shown that a larger ergoregion is accompanied by an increase in superradiance \cite{deOliveira:2020lzp,Liu:2024qso}. However, the ergoregion fails to account for the effect of the parameter $k$ on superradiance in the context of Kerr black-bounce spacetime. The Penrose process in the vicinity of a Kerr black-bounce spacetime was analyzed by Franzin \textit{et al.}~\cite{Franzin:2022iai} to build intuition for the relevant physics. They found that less energy is extracted from Kerr black-bounce spacetimes compared to Kerr spacetimes. This is consistent with the effect of the parameter $k$ on superradiance. Therefore, in the next subsection, we will analyze the parameter dependence through the  EPP to aid our understanding of the effect of parameters on superradiance.

	\subsection{The electric Penrose process around KN black-bounce spacetime} \label{subsec:PP} 
	
	Bhat \textit{et al.} \cite{Bhat:1985hpc} investigated the energy extraction efficiency in the context of a KN BH through the EPP.  They showed that there is no upper limit to the energy extraction when the process of this BH interacting with a charged particle is considered.  Numerous studies have been devoted to investigating this efficiency when considering the electromagnetic interactions \cite{Shaymatov:2022eyz,Viththani:2024map,Alloqulov:2024cto,Xamidov:2025pln}.  
	Consequently, detailed derivations are omitted here, and the efficiency expression is provided directly (see Ref. \cite{Shaymatov:2022eyz} for more details). When a particle with parameters of $\left(q_1, \mu_1, E_1 \right)$ enters the ergosphere from infinity and splits into two particles, a particle falling into the BH  with parameters of $\left(q_2, \mu_2, E_2 \right)$, and the other flying to infinity with parameters of $\left(q_3, \mu_3, E_3 \right)$, satisfying  $q_1\geq q_2+q_3$, $\mu_1\geq \mu_2+\mu_3$ and $E_1=E_2+E_3$, the efficiency expression is
	\begin{align}\label{eq:efficiency}
		\eta_\text{EPP}&=\frac{E_{3}}{E_{1}}-1\\ \nonumber
		&=\left(1+\frac{q_1 A_{t}}{E_1}\right)\left[ \left(\frac{\Omega_1-\Omega_{2}}{\Omega_{3}-\Omega_{2}}\right)\left(\frac{g_{tt}+\Omega_{3}g_{t\phi}}{g_{tt}+\Omega_{1}g_{t\phi}}\right)-1\right]\\ \nonumber
		&-\frac{q_3 A_{t}}{E_1},
	\end{align}
	where $\Omega_{i} = \dd \phi_{i}/\dd t$ denotes the angular velocity of the $i$th particle ($i=1,2,3$), and each component is represented by
	\begin{align}\label{eq:Omega}
		\Omega_{1}&=\frac{-g_{t\phi}\left(u^2+g_{tt}\right)+u \left(\sqrt{-\psi(u^2+g_{tt})}\right)}{g_{t\phi}^{2}+u^2g_{\phi\phi}}, \\
		\Omega_{2}&=\frac{1}{g_{\phi\phi}}\left(-g_{t\phi}-\sqrt{-\psi}\right),	\\
		\Omega_{3}&=\frac{1}{g_{\phi\phi}}\left(-g_{t\phi}+\sqrt{-\psi}\right),
	\end{align}
	with $u=\left(E+q A_{t}\right)/\mu$ and $\psi=g_{tt}g_{\phi\phi}-g_{t\phi}^{2}$.  
	
	It is clear from Eq. \eqref{eq:efficiency} that when $q_1 = q_3 = 0$, i.e., there is no electromagnetic interaction, the particle can only extract the rotational energy of the rotating BH. Bhat \textit{et al.} \cite{Bhat:1985hpc} have demonstrated that in this case, the BH charge $Q$ reduces the maximum efficiency of the Penrose process, while the rotation parameter $a$ has the opposite effect. This behavior is consistent with numerical results on superradiance \cite{Leite:2017hkm}, where superradiance is suppressed by the BH charge $Q$ but enhanced by the spin parameter $a$. Notably, these analyses assume that the test wave or particle propagates along the equatorial plane, $\theta = \pi/2$. When there is electromagnetic interaction around a BH, $A_{t}$ reduces to $\left( Q\sqrt{r^2+ k^2}\right)/\left(r^2+ k^2\right)$ for $\theta = \pi/2$. Therefore, the presence of electromagnetic interactions does not affect the fact that the faster the BH spins, the more energy is extracted by the Penrose process. Benone and Crispino \cite{Benone:2019all} have shown that when charged massive scalar waves propagate along the equatorial plane, the superradiance increases with increasing spin parameter $a$.

	Considering a test particle propagating along the axis, $\theta=0$, we have  $ g_{t\phi}=g_{\phi\phi}=0$ and thus $\Omega_{i}=0$. Therefore, the efficiency $\eta_\text{EPP}$ reduces to 
	\begin{align}\label{eq:efficiency_2}
		\eta_\text{EPP}=-\frac{q_3 A_{t}}{E_1}-1=\frac{q_3 Q\sqrt{r^2+ k^2}}{E_1\left(r^2+ k^2+a^2\right)}-1,	
	\end{align}
	where we can use $E_1 = \mu_1 = \mu$ to estimate the energy of the falling particle for qualitative analysis. Evidently, the energy extraction efficiency is suppressed by the parameters $a$, $k$, and $\mu$. The effect of the field mass $\mu$ on energy extraction efficiency is consistent with its effect on superradiance \cite{Benone:2015bst,Li:2024xyu}. In addition, Franzin \textit{et al.} \cite{Franzin:2022iai} investigated the effect of the parameter $k$ on the Penrose process and superradiance when neutral scalar particles or waves are scattered by the Kerr black-bounce spacetime, with the test particle or wave lying on the equatorial plane. They found that the energy extraction and superradiance are suppressed by the parameter $k$. Xia \textit{et al.}~\cite{Xia:2023zlf} showed that the introduction of other regularization parameters leads to higher energy extraction and stronger superradiance in the context of a rotating loop quantum gravity BH. Based on the above analysis, we can predict that superradiance will be suppressed by the spin parameter $a$ when the plane wave is incident along the rotation axis. 
	
	In Fig. \ref{fig:sur_am_aqk} of Sec. \ref{subsec:superradiance}, we analyzed the effects of the parameters $a$, $q$, and $k$ on the amplification factor (i.e., superradiance), and found that $a$ and $k$ suppress superradiance, while $q$ enhances it. This finding aligns with the analytical results of the current subsection.

	\subsection{Maximal energy extraction and possible application} \label{subsec:AP} 
	
	As an extension of the last subsection, we discuss the maximum extracted energy of the escaping particle when the falling particle is neutral ($q_1=0$), as well as its potential astrophysical application. When $q_1=0$, Eq. \eqref{eq:efficiency} reduces to
		\begin{align}\label{eq:efficiency_3}
			\eta_\text{EPP}
			&= \left(\frac{\Omega_1-\Omega_{2}}{\Omega_{3}-\Omega_{2}}\right)\left(\frac{g_{tt}+\Omega_{3}g_{t\phi}}{g_{tt}+\Omega_{1}g_{t\phi}}\right)-1-\frac{q_3 A_{t}}{E_1}\\ \nonumber
			&=\eta_{\text{PP}}-\frac{q_3 A_{t}}{E_1},
		\end{align}	
		where clearly $\eta_{\text{PP}}$ is the efficiency of the Penrose process affected only by purely geometric factors. When the split point is located at the event horizon \eqref{eq:rh}, the efficiency $\eta_\text{EPP}$ reaches its maximum.  Patel \textit{et al.}~\cite{Patel:2022jbk} have shown that for a Kerr black-bounce spacetime, the maximum value of $\eta_{\text{PP}}$ depends only on the spin parameter $a$, indicating that the parameter $k$ does not affect it. We find that the same phenomenon occurs in a KN black-bounce spacetime due to the simple relation $r_{h}^{\text{KN}} = \sqrt{r_{h}^2 + k^2}$, so that the maximum value of $\eta_{\text{PP}}$ in the KN black-bounce spacetime coincides with that in the standard KN spacetime. Additionally, from Eqs.~\eqref{eq:rh} and \eqref{eq:efficiency_2}, we observe that the term $-q_3 A_{t}/E_1$ is also independent of $k$ at $r = r_{h}$. Therefore, the maximum efficiency of $\eta_\text{EPP}$, realized at the horizon, is independent of the parameter $k$.
	
	For the two contributions to $\eta_{\text{EPP}}$ in Eq. \eqref{eq:efficiency_3}, it is known that the electric part dominates the geometric term $\eta_{\text{PP}}$ when $\eta_\text{EPP}$ reaches its maximum $\eta_\text{ultra}$ at the horizon \cite{Tursunov:2020juz,Tursunov:2021jjf}. Thus, $\eta_\text{EPP}$ can be approximated as 
		\begin{align}\label{eq:efficiency_4}
			\eta_{\text{ultra}}\approx -\frac{q_3 A_{t}}{E_1}=\frac{q_3 Q\sqrt{r_h^2+ k^2}}{E_1\left(r_h^2+ k^2+a^2 \cos^2\theta\right)}.	
		\end{align}
		According to Eq. \eqref{eq:efficiency} and the above expression, the extracted energy $E_3$ of the escaping particle can be expressed as
		\begin{align}\label{eq:E_3} 
			E_{3}^{\text{ultra}}=(\eta_{\text{ultra}}+1)E_1, 
		\end{align}
		which can reach extremely high values. To evaluate $\eta_{\text{ultra}}$, we assume that the escaped particle, modeled as an ion with mass number $A$ and atomic number $Z$, has negligible kinetic energy. Therefore, Eq.~\eqref{eq:efficiency_4} can be written as
		\begin{align}\label{eq:ultra}
			\eta_{\text{ultra}} \approx  \frac{Z e Q r_{\text{h}}^{\text{KN}}}{A m_n c^2\left((r_{\text{h}}^{\text{KN}})^2 + a^2 \cos^2\theta\right)},
		\end{align}
		with $e$ and $m_n$ being the elementary charge and nucleon mass. Regarding the charge $Q$ carried by the black hole, it is generally assumed in astronomy that the black hole rarely reaches its maximal charge due to selective accretion of opposite charges. Therefore, we adopt a conservative value for $Q$ \cite{Tursunov:2021jjf,Juraev:2024dju}
		\begin{align}\label{eq:Range}
			10^{11}\frac{M}{M_{\odot}}\, \text{Fr}\lesssim Q\lesssim10^{18}\frac{M}{M_{\odot}}\, \text{Fr},\quad 1\,\text{Fr}\simeq0.33\times 10^{-9}\text{C}. 
		\end{align}

	\begin{figure}[htp!]
		\centering
		\includegraphics[width=0.48\textwidth]{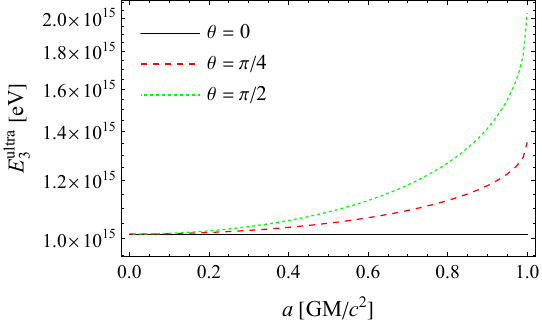}
		\caption{The energy of escape particle $ E_{3}^{\text{ultra}}$ as a function of the spin parameter $a$ for $\theta = 0,\, \pi/4,\, \pi/2$.
		}\label{fig:E_3}
	\end{figure}
	
	Taking the upper limit of Eq.~\eqref{eq:ultra}, we set $Z=A$. We then plot the energy $E_{3}^{\text{ultra}}$ of the escaped particle per nucleon mass (i.e., further setting $Z=1$ for a proton) using Eq.~\eqref{eq:E_3} as a function of the spin parameter $a$ in Fig.~\ref{fig:E_3} for $\theta = 0,\, \pi/4,\, \pi/2$. The value of $Q$ used here is the upper limit of Eq.~\eqref{eq:Range} and is chosen to obtain the optimal extraction energy. We find that $E_{3}^{\text{ultra}}$ increases with $a$ for a test particle along nonzero $\theta$. Moreover, the closer the particle's motion is to the equatorial plane (i.e., the closer $\theta$ is to $\pi/2$), the faster the extracted energy increases with the spacetime spin. For an extreme black hole, the extracted energy along the equatorial direction is exactly twice that along the spin axis. This dependence of the extracted energy on both the spin $a$ and the BH rotation axis orientation $\theta$ provides a potential means to constrain these parameters by measuring the escaped particle energy through the EPP. 
	
	Indeed, previous research has used EPP-accelerated particles in Schwarzschild spacetime to explain ultrahigh-energy cosmic rays (UHECRs) and the knee structure observed in the cosmic-ray spectrum from Sgr A* \cite{Tursunov:2021jjf}. In our case, if we set $a=0$ and use the upper limit of $Q$ in Eq.~\eqref{eq:Range}, we find that
		\begin{align}\label{eq:ultra_1}
			E_{\text{ultra}} = \frac{e Q}{ r_{\text{h}}^{\text{RN}}} \approx 1.015 \times 10^{15}\,\text{eV}.
		\end{align}
	This corresponds to the value of $E_{3}^{\text{ultra}}$ at the left end of Fig.~\ref{fig:E_3} and agrees with what can be deduced from Ref.~\cite{Tursunov:2021jjf}. If the spin $a=M$, the extracted energy equals the previously mentioned value for $\theta=0$, but will be doubled if $\theta=\pi$, i.e., $E_{3}^{\text{ultra}} \approx 2.03 \times 10^{15}\,\text{eV}$, as seen from the right end of the green curve in Fig.~\ref{fig:E_3}. 
	
	\begin{figure}[htp!]
		\centering
		\includegraphics[width=0.48 \textwidth]{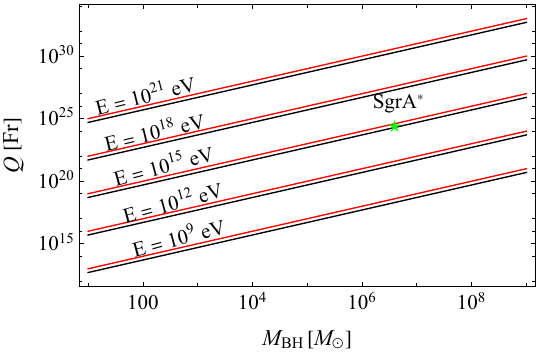}
		\caption{Energy levels of the accelerated protons as functions of $M_\text{BH}$ and $Q$. Red and black lines correspond to the $a=0$ case, or equivalently $(a=M,\,\theta = 0)$, and to the $(a=M,\, \theta = \pi/2)$ case, respectively. The green star marks the theoretical acceleration upper bound of Sgr A*, using the mass and charge of Sgr A* taken from Ref. \cite{Zajacek:2018ycb}.
		}\label{fig:SgrA}
	\end{figure}
	In Fig.~\ref{fig:SgrA}, we further show how to constrain the mass and charge of BHs acting as accelerators for high-energy protons. This figure plots the energy levels of such protons as $M$ and $Q$ change for several $(a, \theta)$ combinations. We find that, regardless of whether a BH rotates, increasing its charge $Q$ for a given mass $M$, or increasing its mass $M$ for a given charge $Q$, can always accelerate the particles to higher energies. Compared to the cases with $a=0$ or $(a=M, \theta=0)$ (red line), a rotating BH can accelerate particles along the equatorial plane $(a=M,\, \theta=\pi/2)$ (black line) with slightly less BH charge to reach the same energy level. Moreover, the green star in the plot shows the acceleration capability of Sgr A* when its charge is maximal (taken from Ref.~\cite{Zajacek:2018ycb}). Clearly, for $(a=M,\,\theta=\pi/2)$, Sgr A* can accelerate ionized particles to energies of the order of $1.82\times10^{15}\,\text{eV}$, which corresponds to the knee of the cosmic-ray spectrum at $\sim 10^{15.5}\,\text{eV}$ \cite{Tursunov:2020juz}. If applied to other BH sources of cosmic rays (provided their spectra are distinguishable from others), this approach is also expected to put constraints on the parameters of these sources.

It is also worth commenting on the detectability of these particles. Ionized particles accelerated to ultrahigh energies via the EPP near supermassive black holes (SMBHs)—such as protons reaching energies of order $\sim 10^{15}$~eV or even higher—could be detected as extensive air showers by ground-based observatories such as the Pierre Auger Observatory (PAO) and the upcoming Cherenkov Telescope Array (CTA). The former has been measuring UHECRs (energies $\gtrsim 10^{18}$~eV) since 2004 \cite{PierreAuger:2007pcg} and excels at determining their arrival directions, composition, and fluxes to trace galactic origins \cite{Boncioli:2014kla,PierreAuger:2017pzq,PierreAuger:2020kuy}. Using Eqs.~\eqref{eq:E_3} and~\eqref{eq:Range}, and referring to Fig.~\ref{fig:E_3} (for Sgr A*), we find that SMBHs with masses larger than $\sim 10^9 M_\odot$ can generate EPP-accelerated UHECRs with energies above $10^{18}$~eV, which fall within the detectable range of PAO. Such SMBHs include well-known examples such as M87*.  Complementarily, the upcoming CTA will detect Cherenkov radiation from air showers induced by charged cosmic rays and gamma rays, offering high-resolution imaging and energy reconstruction down to $\sim 10^{12}$~eV \cite{CTAConsortium:2010umy,CTAConsortium:2017dvg}. For Sgr A*, if the total central charge $Q$ is below the upper limit given in Eq.~\eqref{eq:Range}, the energies of accelerated charged particles (from Eq.~\eqref{eq:E_3}) could be several orders of magnitude lower than those shown in Fig.~\ref{fig:E_3}, thereby falling well within the detection range of the CTA. Therefore, such detectors may reveal anisotropic UHECR fluxes aligned with known galactic centers, enabling tests of our model predictions on energy-extraction efficiency and constraints on black-hole parameters.

	\subsection{Superradiance on scattering cross section} \label{subsec:sur_sec}
	
	\begin{figure}[htp!]
		\centering
		\includegraphics[width=0.48 \textwidth]{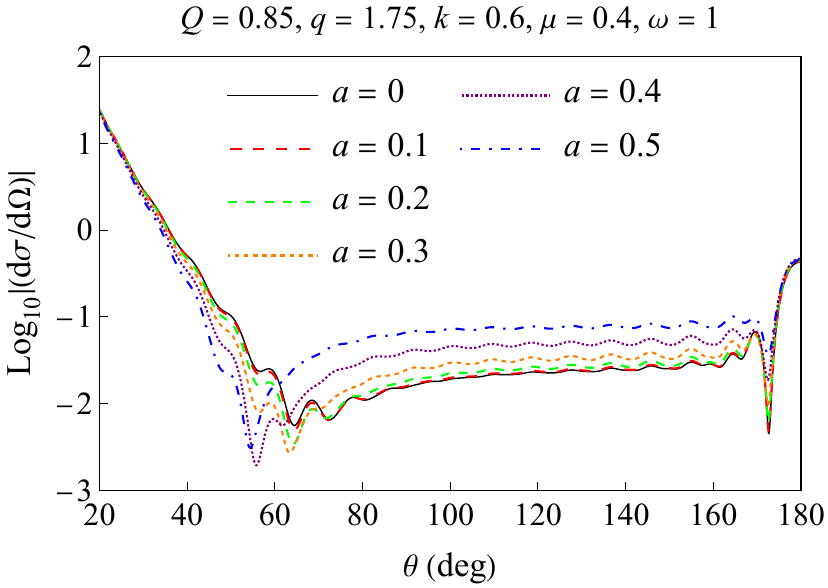}
		\includegraphics[width=0.48 \textwidth]{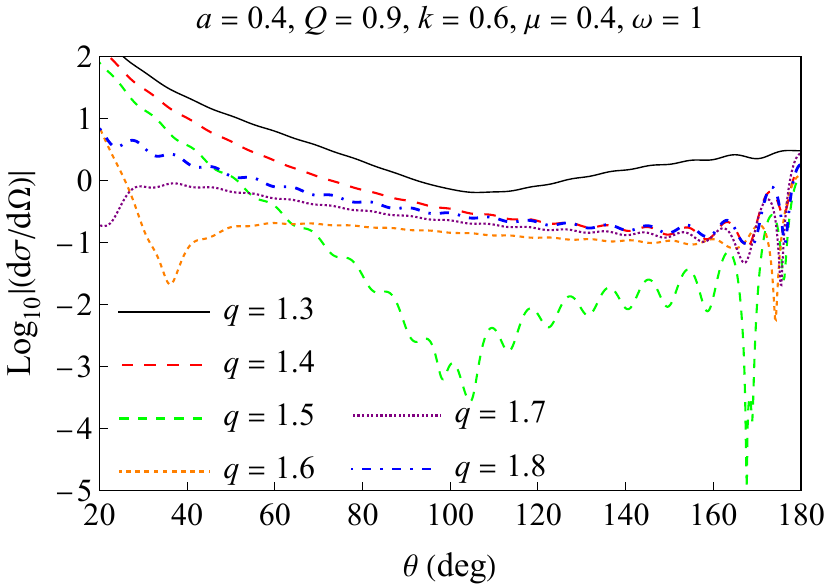}
		\includegraphics[width=0.48\textwidth]{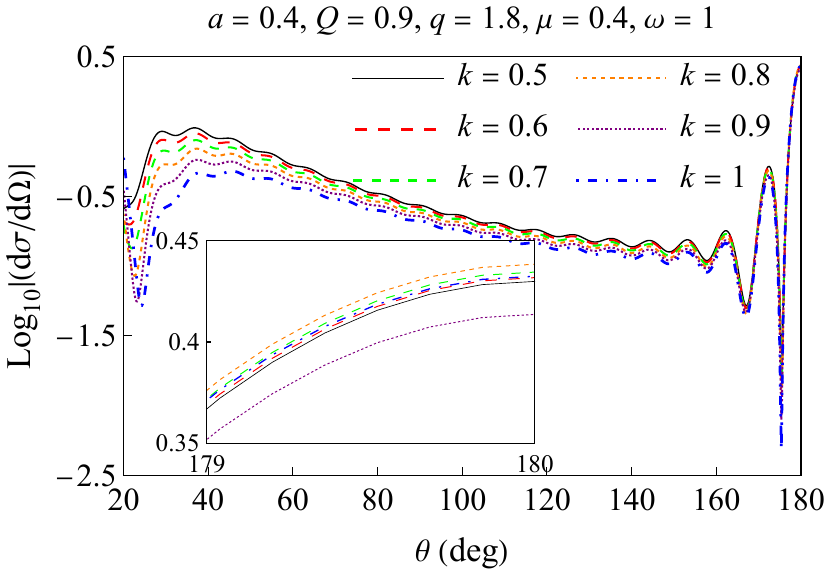}
		\caption{The differential scattering cross section for different values of $a$ (top), $q$ (center) and $k$ (bottom). The inset is the differential scattering cross section near the scattering angle $\theta \sim \pi$ for the different values of $k$.  }\label{fig:sur_sca}
	\end{figure}
	Turning our attention to the variation of the differential scattering cross section under superradiance conditions, we observe from the bottom left panel in Fig. \ref{fig:sca1} that as the Lorentz repulsive force intensifies, the differential scattering cross section decreases while its interference fringes broaden. Concurrently, we anticipate an amplification in the scattered flux intensity due to the occurrence of superradiance.  To systematically compare the effects of these three parameters, in Fig. \ref{fig:sur_sca}, we vary the values of $a$ (top), $q$ (center), and $k$ (bottom). Notably, variations in $a$ (top) and $q$ (center) cause $\omega_c$ to cross a fixed value of $\omega$, enabling us to observe the resulting variability in the differential scattering cross section.
	
	From the top plot, it is evident that for any fixed $a$, the differential scattering cross section first decreases as $\theta$ increases from small values until some intermediate value around $60^\circ$ and then keeps increasing from there. This is slightly different from the top left plot of Fig. \ref{fig:sca1}, primarily due to different choices of the parameters. For increasing $a$ but fixed $\theta$, we see that the cross section also decreases at small $\theta$ while increasing at larger $\theta$ although the absolute values at larger $\theta$ are much smaller than the values at small angles.  Before and after the superradiance happens (roughly when $a> a_c\approx 0.33$), we do not see any apparent qualitative change in this cross section besides the fact that it changes smoothly as $a$ passes through its critical value. 
	
	From the center plot, we see that the cross section in the whole angle range declines as $q$ increases until it reaches the superradiant value of $q_c\approx 1.56$, where the cross section at large angles starts to increase as $q$ increases.     Moreover, as $q$ approaches $q_c$, a dip in the cross section at some intermediate angle starts to appear and this dip as $q$ increases becomes weaker and shifts toward smaller angles. 
	
	Finally, for the effect of $k$, since it does not affect $\omega_c$, we can only show the differential scattering cross section when superradiance has occurred $(\omega=1<\omega_c)$ in the bottom plot of Fig. \ref{fig:sur_sca}. We see that as $k$ increases the differential scattering cross section decreases smoothly for almost all $\theta$ except for very large or small angles around $\theta=\pi$ or $\sim 20^\circ-\sim 30^\circ$. At these limit angles, the oscillating feature of the cross section becomes apparent, and this causes the order of the cross section to mix due to the squeezing/widening effect of $k$ on the oscillation peaks.

	\section{Conclusions}\label{sec:conclusion}
	In this work, the absorption and scattering cross sections of the KN black-bounce spacetime for a charged massive scalar wave propagating along the rotation axis are studied. In studying geodesic absorption and scattering, we presented the first investigation of charged massive particle motion in KN black-bounce spacetime. Notably, we found identical critical impact parameters for both KN and KN black-bounce spacetimes.  The geometrical cross section of absorption, the glory and classical differential scattering cross sections and the corresponding numerical results obtained by the partial wave method were compared for varying spacetime parameters ($a,\,Q,\,k$) and field parameters ($q,\,\mu$) as well as the kinetic variables $\omega$ and $\theta$. We paid special attention to the effects of the rotating parameter $a$, electromagnetic interaction and the regularization parameter $k$ on these cross sections when superradiance happens. 	To further explore the physics of superradiance, that is, why it exhibits such behavior, we briefly studied the EPP in this spacetime. As an extension of the EPP, we discussed the maximum energy of ionized particles and its potential astrophysical application, namely, that the EPP could provide a possible explanation for why SMBHs may serve as sources of UHECRs, as well as how such particles, when observed, could be used to constrain BH spacetime parameters.

	The main finding is that, in general, a faster (or slower) rotating spacetime or a more repulsive (or attractive) electric force tends to reduce (or increase) the absorption cross section and cause wider (or narrower) interference fringes in the scattering waves.  We noticed that in the case of on-axis incidence, a slower rotating spacetime can lead to an increase in the extent of superradiance, as opposed to the case of equatorial incidence in a KN BH \cite{Benone:2019all}. This behavior is consistent with the influence of the parameter $a$ on the energy extraction efficiency of the EPP, in which a larger spin suppresses the efficiency for particles moving along the rotation axis.  When the repulsive interaction is strong enough, the absorption cross section is finite even when $\omega\to\mu$. Moreover, the presence of stronger repulsive electric field forces in the surrounding spacetime would further increase the extent of superradiance. However, the peak of $\sigma_{00}$ does not increase with the Lorentz repulsion like the peak of $Z_{\omega 00}$, but it increases and then decreases.  Therefore, the intensity of superradiance cannot be determined simply from the peak of $\sigma_{00}$.  The parameter $k$ only weakly modifies the absorption/scattering cross sections, but can noticeably suppress superradiance. This suppression effect is consistent with the results of the neutral scalar wave scattered by a  Kerr black-bounce spacetime \cite{Franzin:2022iai}. For the effect of field mass, it is found that a heavier scalar field is more easily absorbed and its corresponding differential scattering cross section exhibits wider interference fringes.  When superradiance happens, we do not see any qualitative change in the differential scattering cross section, such as a case where the absorption cross section becomes negative.

	When the test field has a mass, as in the case studied here, it was known that superradiant instabilities may be excited. Moreover, as shown previously \cite{Huang:2018qdl} and demonstrated in the present work, electromagnetic interactions can lead not only to a higher intensity of superradiance, but can also make it easier to reach the superradiant instability. This fact has important astrophysical implications. It plays a key role in the discovery of new hairy BH solutions, which can circumvent classical no-hair theorems \cite{Herdeiro:2014goa}. In addition, superradiant instability can turn astrophysical BHs into effective particle detectors for ultralight fundamental fields \cite{Arvanitaki:2009fg}, which is important for understanding the nature of dark matter. For further discussions on this phenomenon and its astrophysical applications, we refer the reader to the comprehensive review \cite{Brito:2015oca}. 
Finally, let us also discuss the possibility of constraining the parameter $k$ using gravitational waves (GWs). 
There are fundamentally two ways to do this. The first is when a GW is scattered by a BH with a nonzero $k$, whose signature might be reflected in the cross sections of the scattered GWs when they are detected, although detecting such scattered GWs is extremely challenging with current instruments. A second approach is when one (or both) of the merging objects is a black-bounce BH. In this case, it is natural to expect that during the ringdown phase of the merger, the GWs will show features different from those of standard Kerr BH GWs, because from the metric \eqref{eq:spacetime} it is clear that any signature of $k$ should arise from the modified near-horizon geometry. In our opinion, the second way is more promising, in that the GWs from mergers have been detected and therefore can be directly compared to theoretical predictions for the effects of $k$ once they become available in the future, while the scattering of GWs has not yet been observed. Some preliminary results on the quasinormal modes of scalar fields in black-bounce BH spacetimes can be found in Refs.~\cite{Franzin:2022iai,Santos:2025xbk,Ou:2021efv}.

	\section*{ACKNOWLEDGMENTS}
	One of the authors (Q. Li) thanks for the discussion with Mr. Jinhong He. This work is partially supported by a research development fund from Wuhan University.

\end{document}